\documentclass{aastex631}
\usepackage{amsmath}
\usepackage{hyperref}
\shorttitle{NUV of M dwarfs}
\shortauthors{Jao et al.}
\graphicspath{{./}{figures/}}

\begin{document}
\title{Mind the Gap II: the near-UV fluxes of M dwarfs}

\correspondingauthor{Wei-Chun Jao}
\email{wjao@gsu.edu}

\author[0000-0003-0193-2187]{Wei-Chun Jao}
\affil{Department of Physics and Astronomy \\
Georgia State University \\
Atlanta, GA 30303, USA}

\author[0000-0002-1176-3391]{Allison Youngblood}
\affil{NASA Goddard Space Flight Center, Greenbelt, MD 20771, USA}



\begin{abstract}

Because of the continuous variations in mass, metallicity, and opacity, dwarf stars are distributed along the main sequence on optical and near-IR color-magnitude diagrams following a smooth polynomial. In this study of utilizing a catalog of cross-matched {\it GALEX} and {\it Gaia} sources, we identify two distinct populations of M dwarfs in the near-ultraviolet (NUV) band on the $M_{NUV}$ vs. $M_G$ diagram. We also reveal a pronounced increase in the number of stars exhibiting high NUV fluxes near the spectral type M2 or $M_G\sim9.4$, coinciding with the $H_2$ formation in the atmosphere to improve the energy transportation at the surface. This suggests that certain yet-to-be-understood stellar mechanisms drive heightened activity in the NUV band around the effective temperature of M2 and later types of M dwarfs. Through examination of archival Hubble Space Telescope spectra, we show that Fe II line forests at $\sim$2400\AA~and 2800\AA~dominate the spectral features in the {\it GALEX} NUV bandpass, contributing to the observed excess fluxes at a given mass between the two populations. Additionally, our investigation indicates that fast rotators and young stars likely increase brightness in the NUV band, but not all stars with bright NUV fluxes are fast rotators or young stars.

\end{abstract}

\keywords{M dwarf stars (982) --- Main sequence stars(1000) --- Stellar chromospheres(230) --- Ultraviolet photometry(1740)}


\section{Introduction} 
\label{sec:intro}

The Hertzsprung-Russell diagram (hereafter HRD) is one of the most fundamental figures in stellar astronomy. The locations of stars on this graph reveal and differentiate many aspects of stars, such as temperature, luminosity, binarity, metallicity, and evolutionary stages. Hydrogen-burning dwarfs, spanning from early F and G to late M dwarfs, populate the main sequence on this diagram. The main sequence's continuity arises from the seamless transition of stellar masses, effective temperatures, and opacities across its span. Consequently, polynomial relationships are often employed to elucidate correlations among different stellar parameters, such as color-effective temperature \citep{Casagrande2008}, mass-radius \citep{Parsons2018, Schweitzer2019, Pineda2021}, radius-luminosity \citep{Boyajian2012} or mass-luminosity relations \citep{Benedict2016, Mann2019, Chevalier2023}, particularly for cool dwarfs. These relationships facilitate population studies, including characterizing M dwarfs in surveys like APOGEE \citep{Souto2020} and providing homogeneous stellar parameters for stars in catalogs such as Kepler \citep{Berger2020} and TESS catalogs \citep{Stassun2019}. 

In this study, we examine population distributions in the near-ultraviolet (NUV) band using the {\it GALEX} All-Sky Imaging Survey catalog cross-matched with {\it Gaia} DR2 (``GUVmatch\_AISxGaiaDR2'') from \cite{Bianchi2020}, part of the High-Level Science Products (HLSP) available through the Mikulski Archive for Space Telescopes (MAST). Unlike the aforementioned relations, our analysis shows two distinct polynomial representations to characterize cool dwarfs in the NUV. We discuss our sample selection in section 2 and the connection between the NUV anomaly at M2 and the $H_2$ formation in section 3.
In section 4, we review previous studies addressing these populations. Additionally, we try to establish connections between H$\alpha$, rotation, and NUV fluxes in sections 5 and 6. Sections 7 through 9 discuss quantifying the excess fluxes between these two populations, studying Mg II doublet and NUV flaring from our sample. Finally, we compare NUV spectra between these two populations in section 10 and summarize our findings in section 11.

\section{Sample}
\label{sec:sample}

We select stars in the ``GUVmatch\_AISxGaiaDR2'' catalog that meet the following limits: $\pi>$ 10 mas,
$\pi/\pi_{err}> 10$, $G_{BP}-G_{RP}>$1.5 (hereafter $BP-RP$), and 7.0$<M_G<$14.0, so these stars have high-quality distances and mostly cooler than K dwarfs. Because this catalog was originally generated matching the Gaia DR2 catalog, we then cross-matched this selected sample with the Gaia DR3 catalog \citep{GaiaDR3} to have updated Gaia astrometry and photometry. Also, due to the large pixel size of 1\farcs5 per pixel in the {\it GALEX} mission and pointing errors, we applied two additional selection criteria to reduce the NUV photometry contamination:

\begin{enumerate}
    \item RUWE $<$ 1.4 to minimize binary contamination\footnote{The RUWE (Re-nomralized Unit Weight Error) limit considering for a ``good'' astrometric solution is discussed in the {\it Gaia} Technical Note of GAIA-C3-TN-LU-LL-124-01.}.
    \item no other source in DR3 with $\Delta G_{RP}<$ 4.0, a magnitude difference at {\it Gaia} RP filter, within 15\arcsec~of the target, thereby providing minimal contamination within 10 pixels of the star observed by the {\it GALEX} mission.
\end{enumerate}

\begin{figure}
    \centering
    \includegraphics[scale=0.80]{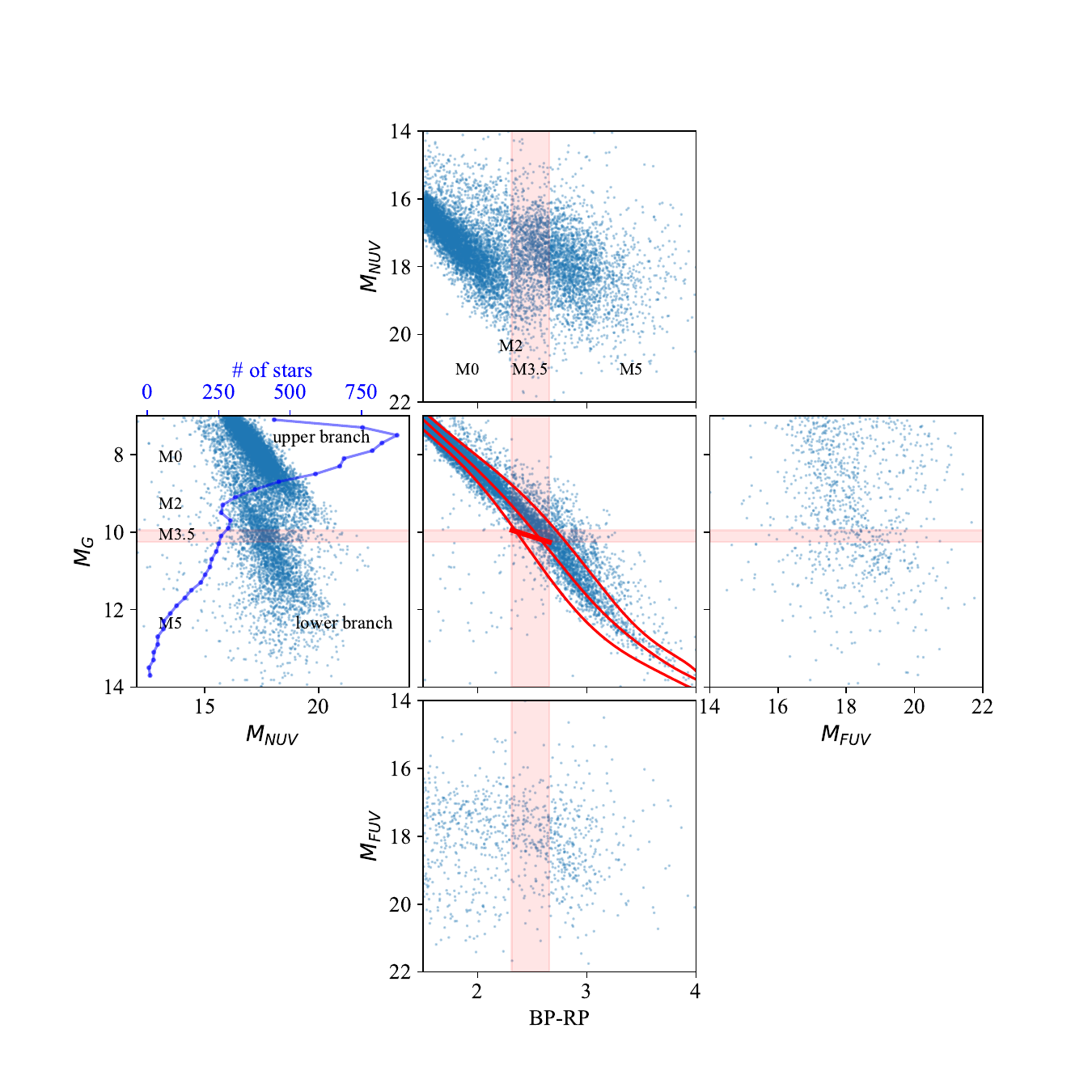}
    \caption{Shared axes plots for the GAGDR3 sample. The center plot is the HRD in $M_G$ vs. $BP-RP$. Three thin red lines mark the general distribution of the main sequence. The top and bottom red lines mark the envelopes of the main sequence, and the center red line is the best-fitted line for the main sequence. The thick angled red line indicates the top edge of the main sequence gap. All these red lines are defined empirically in \cite{Jao2023}. Stars above the gap have a partially convective interior, and stars below the gap are fully convective. The top and bottom plots are the HRD plotted in absolute NUV and FUV magnitudes, respectively.  The left and right plots are the absolute G magnitudes against the absolute NUV and FUV magnitudes, respectively. The vertical and horizontal colored boxes indicate the approximate range of the main sequence gap. Blue dots, connected in a blue line, in the left plot are the number of stars within a bin size of $M_G=$0.2 mags. Three spectral types are labeled at their approximate positions. The two populations of M dwarfs are on self-defined ``upper'' and ``lower'' branches in the left plot, and we will mention these two terms throughout this work.}
    \label{fig:corner}
\end{figure}

In total, 10,234 stars (hereafter GAGDR3) with both {\it GALEX} and {\it Gaia} entries meet the specified criteria. Their distribution on the HRD in $M_G$ vs. $BP-RP$ is shown in the central plot of Figure~\ref{fig:corner}. The majority of these stars align with the main sequence, encompassing dwarfs ranging from late K to approximately M6. However, some stars are positioned above the main sequence, suggesting they may be young or unresolved equal-mass binaries. Conversely, others lie below the main sequence,  potentially indicating they are single subdwarfs or (sub)dwarfs accompanied by unresolved white dwarf companions. It is worth noting that the white dwarf sequence is beyond the range of this plot.

The $M_{NUV}$ vs. $BP-RP$ plot at the top of Figure~\ref{fig:corner} shows a different distribution with a discontinuity. This discontinuity suggests a fragmentation in the main sequence, a phenomenon also apparent in the left plot of $M_G$ vs $M_{NUV}$. Plots on the top and the left indicate two different populations on the main sequence: high and low NUV fluxes observed by {\it {\it GALEX}}. The colored boxes in this figure indicate the range of the main sequence gap first identified by \cite{Jao2018}, where the gap marks the transition between the partial and fully convective interiors. These colored boxes in Figure~\ref{fig:corner} firmly illustrate that the discontinuity lies slightly above the gap, coinciding with a spectral type around M2 or $M_G\sim$9.4, where it aligns with a dip in the distribution of stars within each $M_G$ bin, shown as a thin blue line in Figure~\ref{fig:corner}. This distribution dip is not physical, and we will discuss this dip next. For comparison, similar plots at the FUV band are in the bottom and right of Figure~\ref{fig:corner}. However, because far fewer stars have fluxes detectable in the FUV band and the distributions are sparse, it is difficult to confirm the existence of the two populations in the FUV band. However, generally, most of the stars in the $M_G$ vs $M_{FUV}$ plot are those stars with high NUV fluxes in the $M_G$ vs $M_{NUV}$ plot.

\begin{figure}
    \centering
    \includegraphics[scale=0.7]{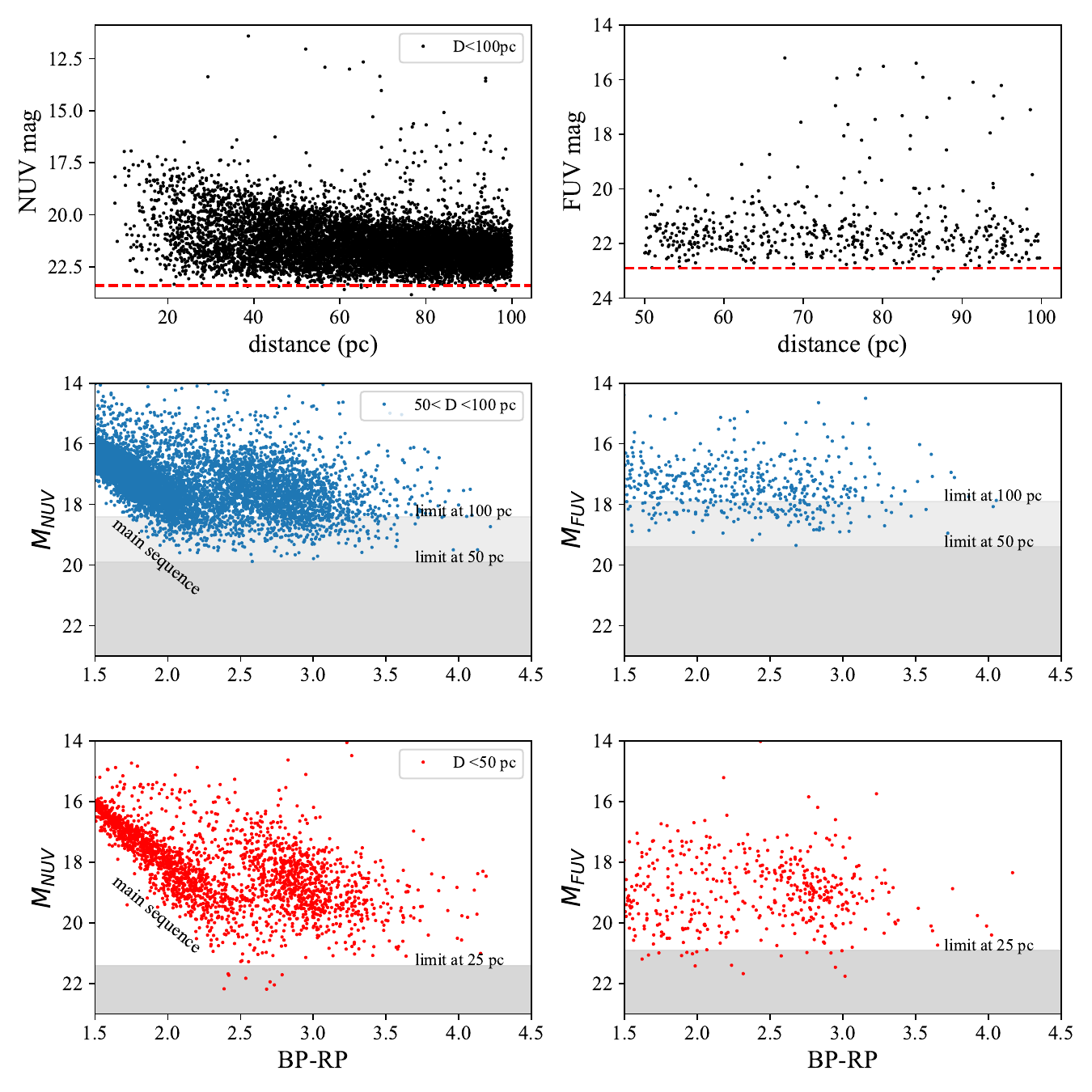}
    \caption{Magnitude limits of {\it GALEX} at various distances. The top two plots show apparent magnitudes in NUV and FUV for our sample (black dots) against distances. Two red dashed lines mark the approximate {\it GALEX} apparent magnitude limits at 23.4 and 22.9 mags for NUV and FUV bands, respectively. The bottom four plots show the absolute magnitude limits in the NUV and FUV bands at various distances marked by shaded gray boxes, and the absolute magnitude limits are given in Table~\ref{tbl:limits}. Blue dots are stars with distances between 50 and 100 pc, and red dots represent stars within 50 pc.}
    \label{fig:HRD}
\end{figure}

The reported apparent magnitude limit in NUV band from the {\it GALEX} mission is $\sim$25.5 mag\footnote{http://www.galex.caltech.edu/researcher/faq.html}. However, as shown in the upper-left panel of Figure~\ref{fig:HRD}, no star in our sample has an apparent NUV magnitude fainter than 24. Hence, we adopted the limits based on our sample in NUV and FUV bands to include 99.98\% of stars. The apparent and absolute magnitude limits in NUV and FUV at various distances are also marked in Figure~\ref{fig:HRD} and given in Table~\ref{tbl:limits}. 
Because of these limiting magnitudes, faint late M dwarfs on the upper branch are not observable by {\it GALEX}, which causes a distribution dip around M2 or $M_G\sim$9.4 in the left plot of Figure~\ref{fig:corner}. Although this work focuses on results in the NUV band, for completeness, we continue to plot the distributions in the FUV band on the right column of Figure~\ref{fig:HRD}, which shows a sparse distribution of the sample at different distances across all M dwarfs.

\begin{deluxetable}{ccc}
\label{tbl:limits}
\tablecaption{Magnitude limits in NUV and FUV bands for stars in GAGDR3}
\tablehead{
\colhead{} & \colhead{NUV} & \colhead{FUV}}
\startdata
 & \multicolumn{2}{c}{apparent magnitude} \\
 &  23.4 & 22.9 \\
\hline 
distance (pc) & \multicolumn{2}{c}{absolute magnitude} \\
100 & 18.4 & 17.9 \\
50 & 19.9 & 19.4 \\
25 & 21.4 & 20.9 \\
\enddata
\end{deluxetable}

\begin{figure}
    \centering
    \includegraphics[scale=0.7]{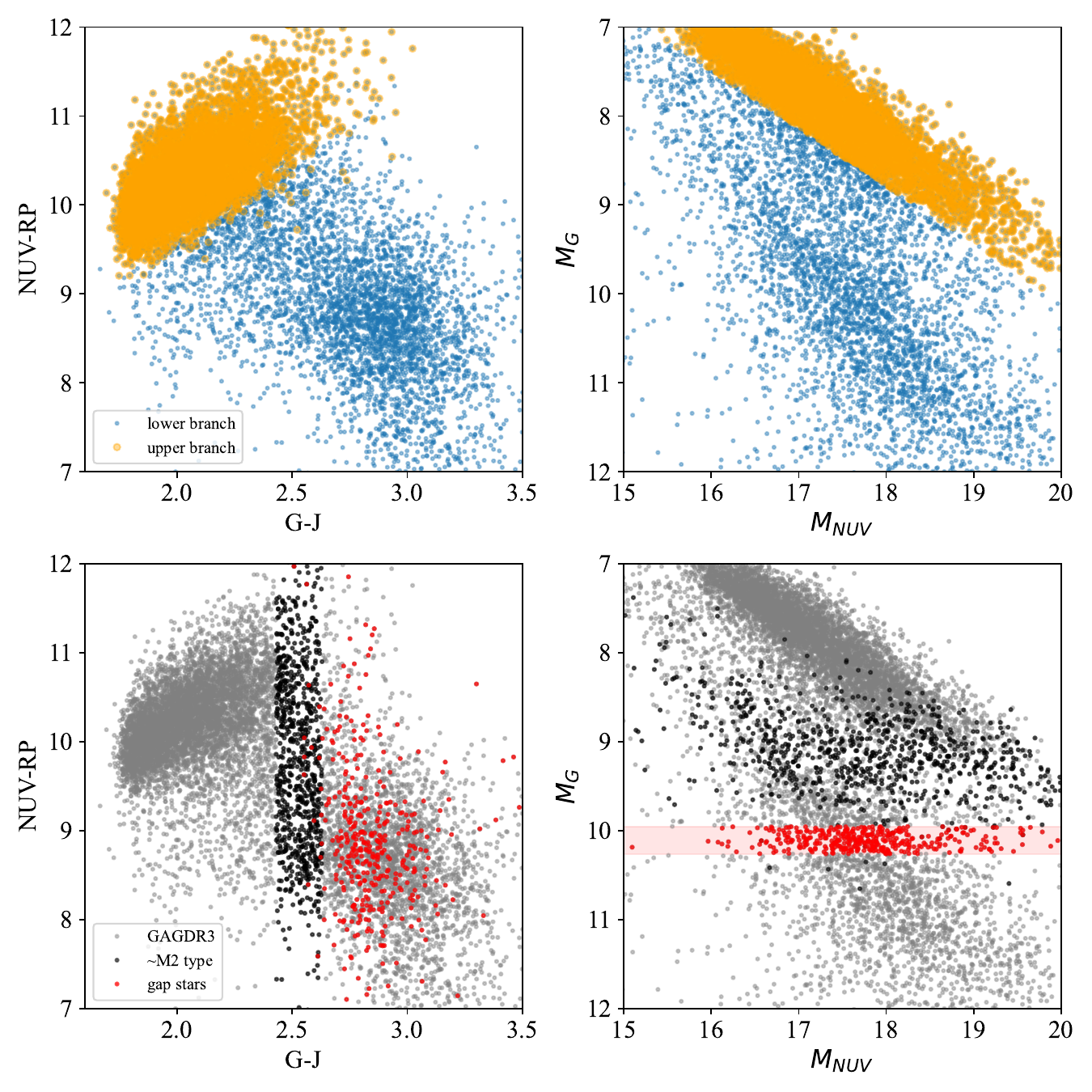}
    \caption{The four-panel figure demonstrates the $NUV-RP$ color shift identified by \cite{Cifuentes2020} at $G-J\sim$2.5 is not at the interior transition based on the GAGDR3 sample. The top two panels show the distributions of stars in $NUV-RP$ vs. $G-J$ and $M_G$ vs. $M_{NUV}$ plots. We also approximately separate stars into upper branch (orange dots) and lower branch (blue dots) using the line later discussed in section~\ref{sec:two_pops} and Table~\ref{tbl:coeff}. The bottom two panels highlight stars with $G-J$ between 2.6 and 2.4 mags at the color shift or close to the M2 type. Red dots represent stars within the main-sequence gap, which is marked by the red rectangle box.}
    \label{fig:CARMENES}
\end{figure}

\section{mass-luminosity relation kink and the M2 anomaly}

\cite{Copeland1970} and \cite{Kroupa2002} discussed the formation of the kinks in the mass-luminosity relation (MLR) using theoretical models. They presented that the first kink close to the M2 type relates to the formation of the $H_2$, and the second kink at M5 is related to the increasing degeneracy pressure in the core, which limits the core contraction \citep{Kroupa2002}. Such kinks can also be identified in the empirical MLR reported in \cite{Benedict2016} and \cite{Mann2019}. The results of the kinks can be seen as the curves of the main sequence highlighted in the red lines in Figure~\ref{fig:corner} (center) near M2 and M5. Here, we only focus on the first kink close to M2.

\cite{Chabrier2000} showed that the collision-induced absorption of $H_2$ and $H_2-He$ provide significant continuum opacity for stars below the effective temperature of 5,000K. For stars below 4,000K, most hydrogen is in the form of $H_2$ in the atmosphere. $H_2$ transports energy very efficiently, causing the onset of convection up to and above the photosphere \citep{Kroupa2002}. Coincidentally, the location of the first kink aligns with the NUV anomaly we found at M2 in Figure~\ref{fig:corner}. We suspect the improved energy transport may carry extra energy into the atmosphere and make these stars brighter in the NUV band.

However, it is not clear why the extra energy is only seen in the NUV band but in neither the optical nor infrared band. In other words, the $M_G$ vs. $M_V$ or $M_G$ vs. $M_{Ks}$ relations are polynomials.  Hence, no single polynomial equation can represent relations such as $M_{NUV}$ vs. color or $M_G$ vs. $M_{NUV}$ for all M dwarfs in Figure~\ref{fig:corner}. Details of obtaining these empirical relations are discussed in section~\ref{sec:two_pops}.


\section{Previous Works}

\cite{Cifuentes2020} reported a $NUV-RP$ color shift near the mid-M dwarfs, using the $NUV-RP$ vs. $G-J$ and $NUV-RP$ vs spectral type diagrams. Because this color shift appears to be close to mid-M in their plots, they proposed that the interior transition from partial to full convection at mid-M dwarfs could cause the shift. However, as shown in Figure~\ref{fig:CARMENES}, our analysis reveals that the interior transition is irrelevant to the identified $NUV-RP$ color shift at around M2. This demonstrates that analyzing stars using both absolute magnitudes and colors provided by {\it Gaia} is crucial to pinpoint the exact population where the shift occurs. 
\cite{Schneider2018} studied UV evolution of mid to late-M dwarfs with {\it GALEX} using known young and field stars. They found that M dwarfs with masses between 0.08 and 0.35$M_\odot$ do not follow the same UV evolutionary trend as early M dwarfs. While they study the ratio of photospheric flux density to the total observed NUV and FUV fluxes as a function of effective temperature and mass, they also found that photospheric contribution in the NUV band is relatively constant for young and field stars with effective temperature between 3400 and 3800K (early M dwarfs). For stars cooler than $\sim$ 3200K, however, they found the photospheric fluxes drop for both young and field stars. Their discovery is similar to what we found here. However, using the high-precision {\it Gaia} astrometry and photometry, we can pinpoint the discontinuity.

\begin{figure}
    \centering
    \includegraphics[scale=0.7]{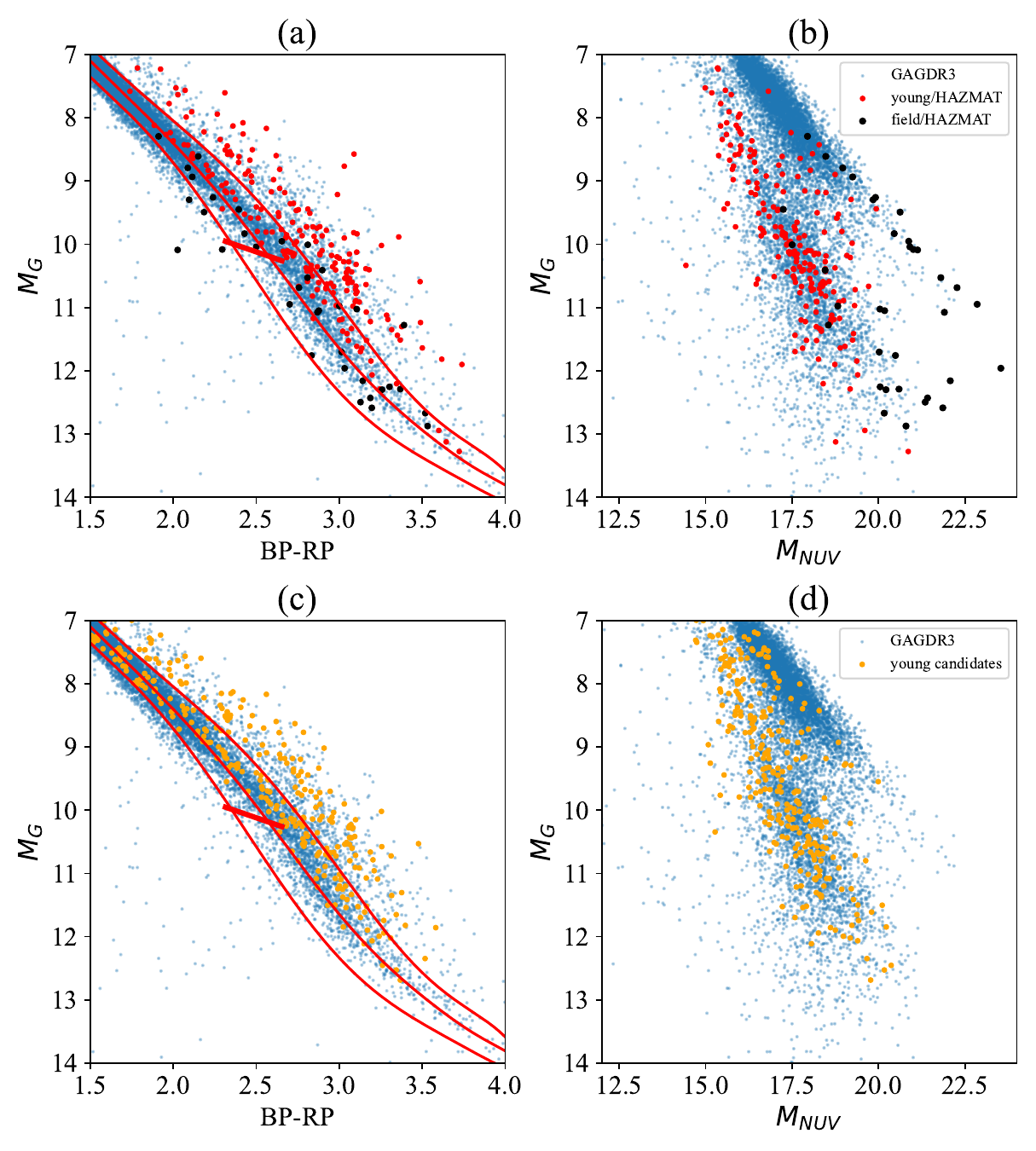}
    \caption{Distributions of stars from GAGDR3 (blue dots) and \cite{Schneider2018} on the HRD (panels a and c) and $M_G$ vs $M_{NUV}$ diagram (panels b and d). Single young and field stars from \cite{Schneider2018} are shown in red and black dots, respectively. Red lines are defined in Figure~\ref{fig:corner}. Stars in the GAGDR3 identified as potential members of young moving groups using the BANYAN tool are shown in orange dots in the bottom two plots.}  
    \label{fig:HAZMAT}
\end{figure}

In \cite{Schneider2018}, they collected NUV and FUV magnitudes of 642 stars from {\it GALEX}, which include field stars and members in moving groups, such as TW Hya, $\beta$ Pic, Tuc-Hor, AB Dor, and Hyades. They also flagged known spectroscopic binaries and visual binaries in their sample. Figure~\ref{fig:HAZMAT} shows their targets on the HRD after applying the same tight selection criteria as for stars from the GAGDR3 discussed in section~\ref{sec:sample}. We remove known binaries and keep single stars flagged in \cite{Schneider2018}, and 249 stars are left in Figure~\ref{fig:HAZMAT}. We note that without vetting all single stars flagged in \cite{Schneider2018}, a few systems may have been resolved since their work. Figure~\ref{fig:HAZMAT} shows most of the flagged ``young'' stars they identified are on the lower branch on the $M_G$ vs $M_{NUV}$ plot. Some of their flagged ``field'' stars are located in the extended region of the upper branch, and some overlap with young stars. The remaining field stars are scattered in the faintest end of the $M_G$ vs $M_{NUV}$ diagram.

Because stars in our sample are all-sky objects, one of the best ways to identify potential memberships among young moving groups is via the BANYAN tool \citep{Gagne2018}. Using astrometry and radial velocities (RV) reported in the {\it Gaia} DR3, 90\% of our stars can provide full kinematics to match memberships. By comparing the radial velocities of 140 M0--M4 dwarfs in \cite{Sperauskas2016} measured by CORAVEL program \citep{Upgren2002}, we found the mean RV difference between the {\it Gaia} and CORAVEL program is merely 0.47$\pm$4.2 km s$^{-1}$. The mean RV errors for the same 140 stars in {\it Gaia} and CORAVEL program are 0.67 and 0.60 km s$^{-1}$, respectively. Therefore, the RV from {\it Gaia} is consistent with the ground-based measurements and can provide reliable stellar kinematics.

We identified potential young stars with BANYAN's membership probabilities greater than 90\%, the same threshold used by \cite{Gagne2018b} to search for new nearby young stars in {\it Gaia} DR2. We found 2.7\% of stars with membership probabilities greater than 90\%, shown at the bottom of Figure~\ref{fig:HAZMAT}. Generally, most young star candidates are in the lower branch of the $M_G$ vs. $M_{NUV}$ plot, similar to those young stars in \cite{Schneider2018}. The remaining stars on the lower branch have low membership probabilities, and we think they
would be 1) young stars not associated with any moving group, 2) rapidly rotating M dwarfs, 3) equal mass short-period binaries, or 4) subject to some yet-to-be-known heating mechanism in the upper stellar atmosphere. Unresolved M and white dwarf binaries could also exhibit high NUV fluxes, but those unresolved binaries would appear to be like cool subdwarfs below the main sequence. As we can see in Figure~\ref{fig:corner}, very few of such systems among GAGDR3 are below the main sequence, so it is unlikely that stars on the lower branch are M$+$ WD binaries. Nonetheless, stars located on the lower branch of the $M_G$ vs $M_{NUV}$ diagram generally, but not all comprise young or active M dwarfs. Old and inactive field stars seem to belong to the upper branch.

\section{The connection between H$\alpha$ activity and NUV photometry}

The H$\alpha$ line at 6563\AA~is a key indicator for stellar activity in the chromosphere \citep{Linsky2017, Basri2022}, and such activities are often connected to stellar rotation and mass \citep{Reiners2014, Wright2018, Newton2017}. In \cite{Jao2023}, they find that the relation between H$\alpha$ activity and rotation exhibits not only mathematically via the activity-rotation relation but geometrically on the HRD: fast rotators and H$\alpha$ active stars are located in the same region of the main-sequence. Essentially, fast rotating and H$\alpha$-active M0--M4 dwarfs are on the top half of the main sequence, and such dwarfs with spectral types of M4 or later are equally distributed on the top and bottom halves of the main sequence. Noticeably, stars in GAGDR3 fainter than $M_G\sim$9 are also mostly elevated to the top half of the main sequence in Figure~\ref{fig:corner}, which resembles the distributions of fast rotating and H$\alpha$-active early M dwarfs described above. This implies close connections between rotation, H$\alpha$ activity, and activity in the NUV band.

\begin{figure}
    \centering
    \includegraphics[scale=0.7]{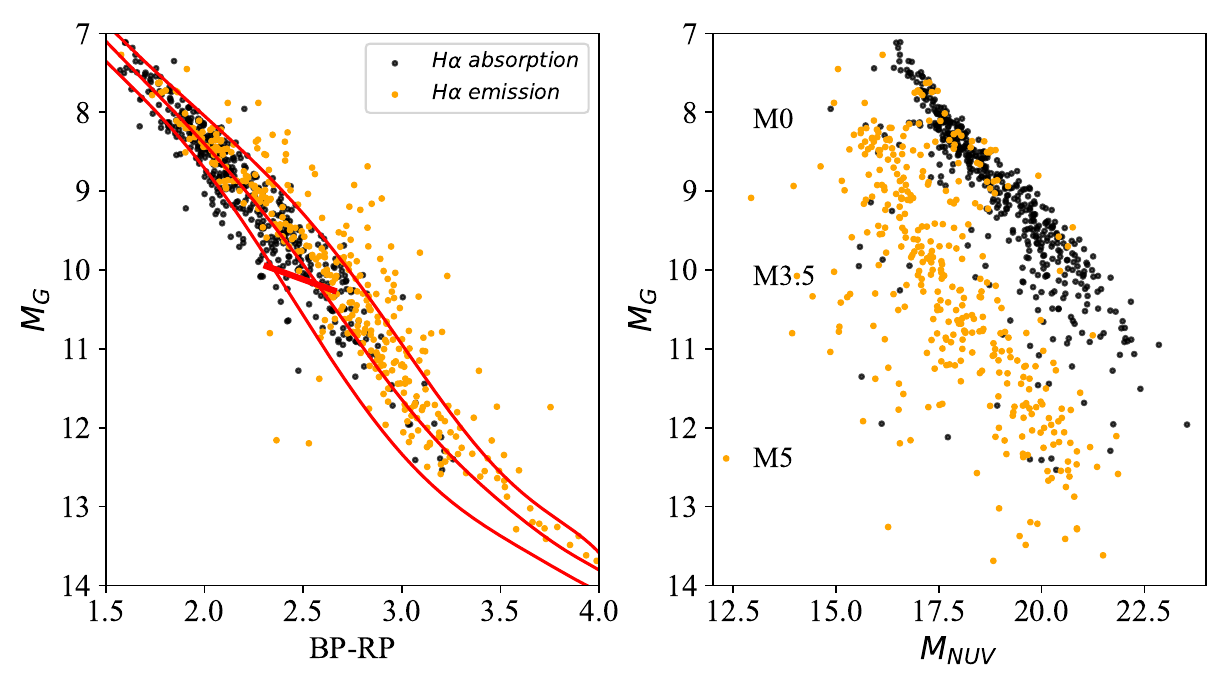}
    \caption{Distribution of H$\alpha$ activity from M dwarfs in the MEarth, CARMENES, and LAMOST surveys. The left plot shows their distributions on the HRD, and the right plot shows their distributions on the $M_G$ vs $M_{NUV}$ plot. The two populations of H$\alpha$ activity mimic the two branches in the right plot. We note that because of the various selection biases from these surveys, the distributions don't indicate the completeness of nearby M dwarfs on these two plots. Red lines on the left plot are defined in Figure~\ref{fig:corner}.} 
    \label{fig:Halpha}
\end{figure}

To further understand the relation between H$\alpha$ activity and NUV fluxes, we collect M dwarfs with H$\alpha$ measurements from \citet[MEarth]{Newton2017}, \citet[CARMENES]{Jeffers2018}, and \citet[LAMOST]{Zhang2021}, and apply the same selection criteria used for sample in the GAGDR3 so that they have the same $BP-RP$ color range, same $M_G$ range, low RUWE, and no field star contamination in {\it GALEX}. Then, this list is cross-matched against the revised catalog of GALEX UV sources (GUVcat\_AIS GR6+7) \citep{Bianchi2017} to obtain their NUV photometry. We found that M dwarfs with H$\alpha$ absorption are mostly on the upper branch in $M_G$ vs. $M_{NUV}$, and M dwarfs with H$\alpha$ emission are on the lower branch. The distributions of these stars are given in Figure~\ref{fig:Halpha}. This further demonstrates that the H$\alpha$ activity in the optical band is strongly related to the fluxes in the NUV band. We note that there are a few subdwarfs below the main sequence showing H$\alpha$ emission, but they are likely to be unresolved close subdwarfs binaries, which increases fluxes in the NUV.

\section{The connection between Rotation and NUV photometry}
\label{sec:rotation}

\begin{figure}
    \centering
    \includegraphics[scale=0.7]{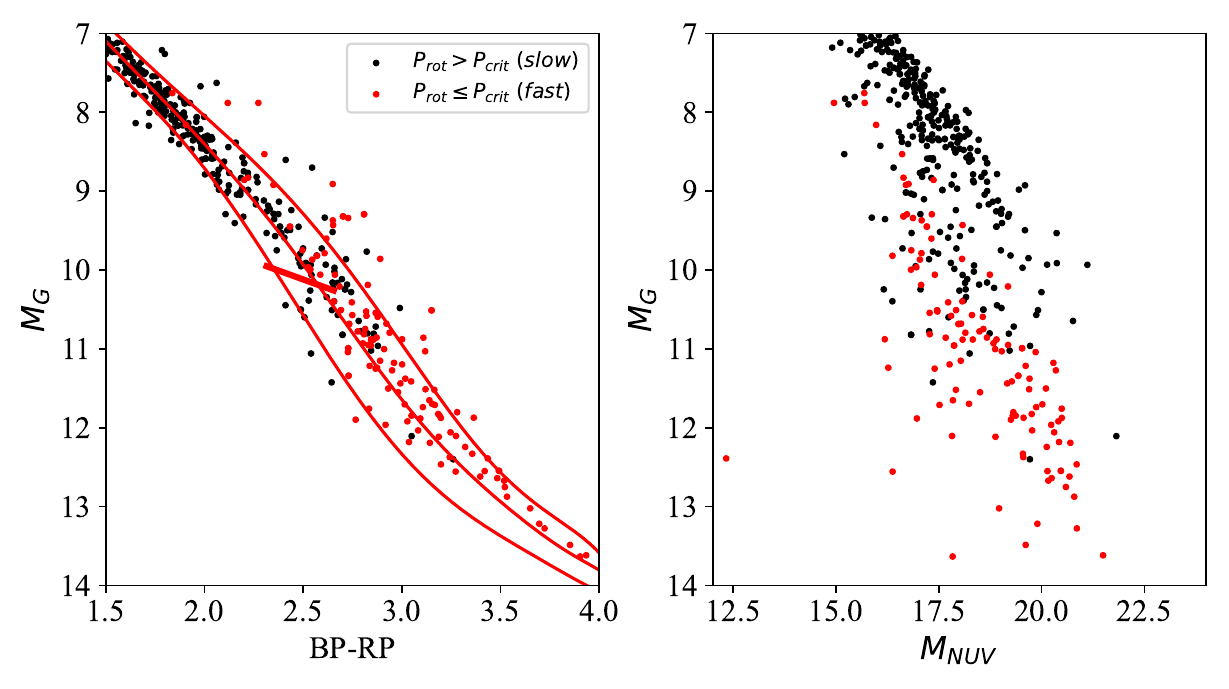}
    \caption{Distributions of stars with rotation period measurement from the Kepler K2 Campaigns in \cite{Reinhold2020} and from nearby stars in \cite{Newton2017}.  The left plot shows their distributions on the HRD, and the right plot shows their distributions on the $M_G$ vs $M_{NUV}$ plot. Black and red dots indicate slow and fast rotators, respectively. The definition of ``fast'' or ``slow'' rotators is relative to the critical period ($P_{crit}$) discussed in \cite{Jao2023}. Red lines on the left plot are defined in Figure~\ref{fig:corner}.}
    \label{fig:rotation}
\end{figure}

Stars with rotation from \cite{Reinhold2020} and \cite{Newton2017} are selected to cross-match the GAGDR3, and all meet the criteria of RUWE$<$1.4 and $\pi>$ 10 mas. We followed the procedures described in \cite{Jao2023} to separate these stars into two groups: fast and slow rotators. Essentially, fast rotators have their rotation periods faster than the critical period ($P_{crit}$) at the kink of the broken power law of the activity-rotation relation and show saturated stellar activity at X-ray or H$\alpha$  \citep{Reiners2014, Newton2017, Magaudda2020}. Slow rotators indicate their rotation periods are slower than this star-dependent critical period, which is determined using the estimated stellar radius and observed rotational period. Our result in Figure~\ref{fig:rotation} shows fast rotators ($P_{rot} \leq P_{crit}$) are all on the lower branch. Conversely, not all stars on the lower branch are fast rotators. This result is similar to Figure~\ref{fig:HAZMAT}, where not all stars on the lower branch belong to young moving groups. \cite{Loyd2021} demonstrated that the strength of NUV and FUV emission lines, such as the Mg II doublet, correlates with the age and rotation of stars and exhibits a similar broken power-law relation to the H$\alpha$ \citep{Newton2017}. Therefore, based on Figure~\ref{fig:HAZMAT} and~\ref{fig:rotation}, most young stars and fast rotators likely appear brighter in the NUV band, but not all stars with high NUV fluxes are young or fast rotators.

\section{Analysis of the two populations}
\label{sec:two_pops}
\subsection{Fitting the Two Populations}

\begin{figure}
    \centering
    \includegraphics[scale=0.7]{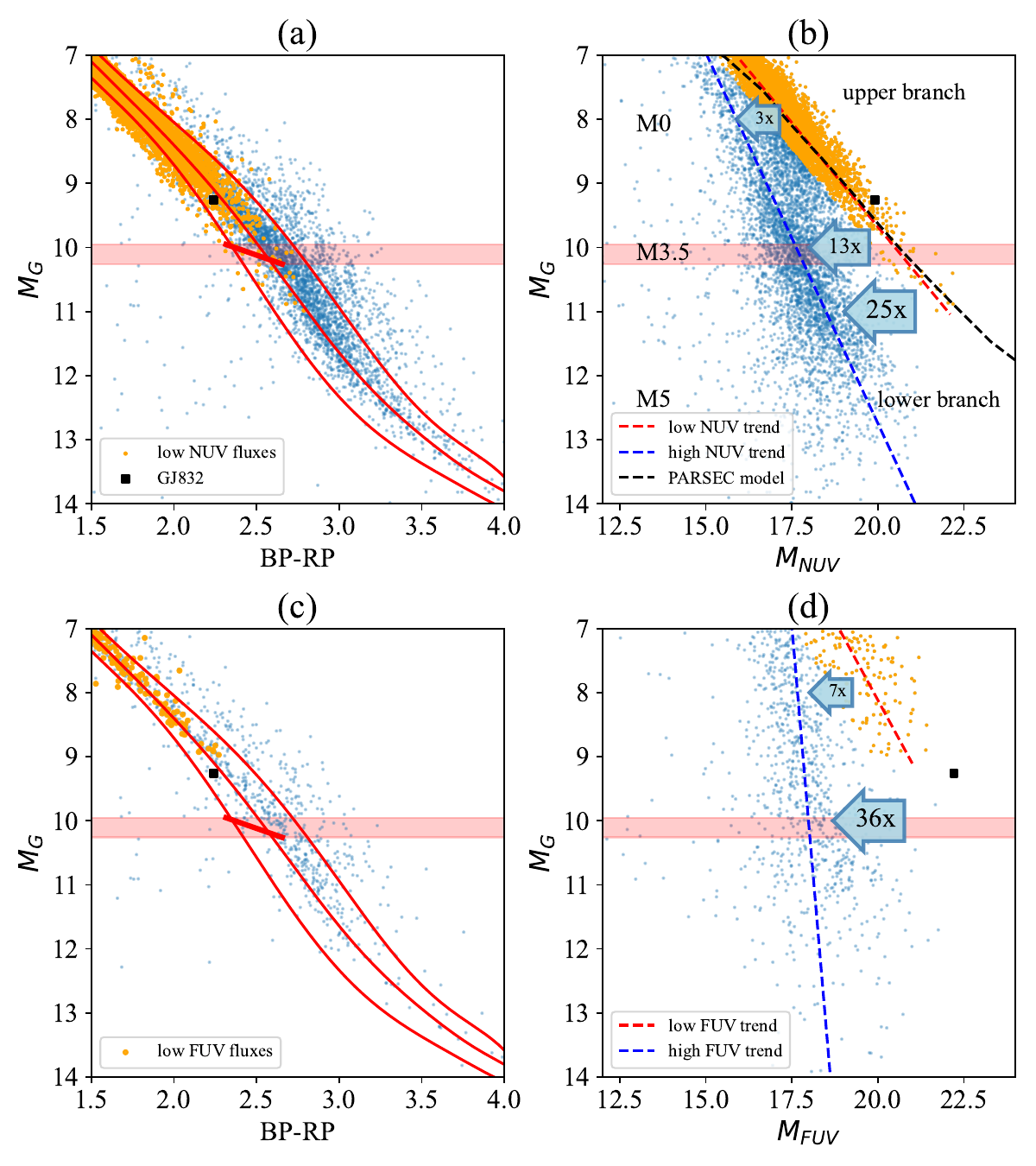}
    \caption{The GAGDR3 sample on the HRD (panels a and c), $M_G$ vs $M_{NUV}$ (panel b), and $M_G$ vs $M_{FUV}$ (panel d) diagrams. Orange dots represent stars on the upper branch, and blue dots are stars on the lower branch. The red and blue dashed lines represent the fitted distributions based on the clustering algorithm. Red lines on panels a and c are defined in Figure~\ref{fig:corner}. A black dashed line in panel b indicates stars with $M_G>$7.0 from the PARSEC model with a 1 billion-year age. The arrows of 3$\times$, 13$\times$, and 25$\times$ represent the excess fluxes relative to the low NUV state at $M_G=8$, 10, and 11, respectively. The bottom two plots are the same as the top two but for stars with detections in the FUV band. The two arrows in the lower right plot indicate the excess fluxes at $M_G=$8 and 10. The PARSEC model in the FUV is not shown here because it is beyond the range of this plot. The NUV benchmark star, GJ 832, is shown as black boxes.}  
    \label{fig:analysis}
\end{figure}

Because there is no distinctive way to separate these two branches of the sample, we apply a clustering algorithm using the Gaussian mixture models from the {\tt scikit-learn} package \citep{scikit} to approximate the two branches. Basically, the clustering algorithm assigns a two-dimensional Gaussian to each distribution to which it most probably belongs. Also, due to the elongated distributions of the two branches, we adopt the full covariance option to classify distributions\footnote{Examples of how to use this package can be found at \\ https://scikit-learn.org/stable/modules/generated/sklearn.mixture.GaussianMixture.html.}. The orientations of the major axes of the two-dimensional Gaussian distributions then represent the linear fit of each branch shown in Figure~\ref{fig:analysis}, and the fitted line coefficients are in Table~\ref{tbl:coeff}. 

Note that we only apply this algorithm to determine the general distribution of each branch, and we don't intend to classify precisely which star belongs to which branch. For plotting purposes, we classify stars as belonging to the upper branch (orange dots in Figure~\ref{fig:analysis}) if they are at most 0.4 magnitudes below the line fitted to the upper branch. This 0.4 magnitude approximately corresponds to the semi-minor axis of the 2D Gaussian distribution assigned by the clustering algorithm. The remaining stars shown in blue are illustrated as stars on the lower branch.

\begin{deluxetable}{ccccc}
\label{tbl:coeff}
\tablecaption{Fitted Distributions in the NUV and FUV bands}
\tablehead{
\colhead{} & 
\multicolumn{2}{c}{NUV} & 
\multicolumn{2}{c}{FUV}}
\startdata
   &  a & b & a & b \\
upper branch & 0.65035 & $-$3.32947 & 6.28459 & $-$103.02273 \\
lower branch & 1.15668 & $-$10.38865 & 0.98846 & $-$11.65192
\enddata
\tablecomments{$M_G=aX+b$, where $X$ is $M_{NUV}$ or $M_{FUV}$.}
\end{deluxetable}

\subsection{Excess Fluxes}
Based on these fitted lines, the magnitude difference between the two branches at a given $M_G$ grows with decreasing masses regardless of the filters. In the NUV band, the difference grows from 1.5 to 3.5 magnitude at $M_G=$8 and 11, respectively. In the FUV band, it grows from 2.2 to 3.9 magnitude at $M_G=$8 and 10, respectively. Because the NUV fluxes are mainly from the photosphere and chromosphere for M dwarfs on the upper and lower branch, respectively, according to models, the different slopes of the lines also mean that for cooler M dwarfs, the chromospheric NUV flux decreases more slowly than the photospheric NUV flux.

Other than presenting the difference between the two branches in magnitude, we can also investigate the ratio of excess flux to the total flux from stars on the upper branch. We assume that stars in the upper branch emit a total flux of $f_t$ from both the photosphere ($f_P$) and the upper atmosphere ($f_C$), with a total magnitude of $NUV_t$. The excess flux, $\Delta f$, contributes additional fluxes to make a star brighter at an ``excited'' or ``high'' state with a magnitude of $NUV_e$.

\begin{eqnarray}
NUV_t=-2.5\times\log_{10}f_t+m_0\\
NUV_e=-2.5\times\log_{10}(f_t+\Delta f)+m_0    
\end{eqnarray}
where $m_0$ is the zero point magnitude and $f_t=f_P+f_C$. Consequently, the magnitude difference is:

\begin{eqnarray}
\Delta NUV = NUV_t-NUV_e = -2.5\log_{10}(\frac{f_t}{f_t+\Delta f})\\
R=10^{-(\Delta NUV/2.5)}\\
\frac{\Delta f}{f_t}=\frac{1-R}{R}
\end{eqnarray}

The ratio of excess to total fluxes,  $\Delta f/f_t$, at a given $M_G$ can be estimated using the fitted trend lines in Figure~\ref{fig:analysis}. We found that the excess flux grows from approximately 3 to 25 times the total fluxes for stars on the upper branch in the NUV band.

The trend line on the upper branch in the NUV band generally matches the isochrone from the PARSEC evolutionary model \citep{Chen2014, Nguyen2022}, which uses the BT-Settl model to set their atmosphere boundary conditions for stars at one billion years old. The isochrone from the PARSEC model is generated using the online 2.0 version and default parameter settings\footnote{http://stev.oapd.inaf.it/cgi-bin/cmd}. Because no upper-atmosphere emission from the chromosphere, transition zone, or corona is included in the PARSEC model, the black dashed line only represents fluxes emitted from the photosphere. This indicates that the upper atmosphere of stars on the upper branch contributes minimally to the NUV band so that the modeled and empirical lines match each other. As for the line comparison between the PARSEC model and the empirical result in the FUV band, they are very different, and the modeled FUV isochrone is beyond this plotting range and far fainter than the empirical line. This is mainly because the chromosphere dominates the emergent FUV fluxes, so the contributions from the photospheres are negligible \citep{Loyd2016}. 

There are far fewer stars with FUV measurements, indicating that most M dwarfs have FUV fluxes fainter than the detection limit of {\it GALEX}. Because of the paucity of stars in FUV, we continue to assume the existence of two populations in the FUV band\footnote{To confirm our assumption, we perform a cursory examination of nearby
K dwarfs on the $M_G$ vs. $M_{FUV}$ distribution. We include K dwarfs
within 60 pc with RUWE$<$1.4, but we do not apply all the same
rigorous selection criteria imposed on the M dwarf sample. We find
that the bulk of K dwarfs in the upper branch are brighter than $M_G$
= 7, and we confirm that stars in the upper branch in Figure 7 (d) are
in the tail of the K dwarf's distribution. Hence, this examination of
including extra K dwarfs confirms the existence of the two populations
in Figure 7 (d). A more rigorous study of K dwarfs and understanding the cause of these two populations are beyond the scope of this work
and will be addressed in our future work. Therefore, we tentatively
only use our sample to fit the two branches in the FUV.}. The red dashed line is fitted by eyes using the bulk of orange dots in Figure~\ref{fig:analysis} to represent the upper branch stars. We continue to use the {\tt scikit-learn} clustering algorithm to fit the lower branch line after the upper branch stars are excluded for fitting. The excess fluxes between these two stages grow from 7 to at least 36 times as the stellar masses decrease. 

Extensive NUV spectroscopic observations for nearby M dwarfs have been conducted from the Hubble Space Telescope (HST) survey programs, such as the HAZMAT (the HAbitable Zones and M dwarf Activity across Time), MUSCLES (the Measurements of the Ultraviolet Spectral Characteristics of Low-mass Exoplanetary Systems), and Mega-MUSCLES (Mega-Measurements of the Ultraviolet Spectral Characteristics of Low-mass Exoplanetary Systems) \citep{Shkolnik2014,France2016,Froning2019}. 
Although the observed trend line that traces the stars in the upper branch matches well with the modeled isochrone, there are differences when comparing individual stellar spectra to the synthetic spectra, especially in the UV. \cite{Loyd2016} compared a PHOENIX photosphere model of GJ 832 and the observed UV spectrum and found fairly good agreement at $\lambda>$2700 \AA, although there are multiple wavelength regions where there is missing emission or absorption in the synthetic spectrum. The model and observed spectra diverge by orders of magnitude at shorter wavelengths, where chromospheric emission dominates over photospheric. Because of its high-quality UV spectra, GJ 832, marked in Figure~\ref{fig:analysis}, is a benchmark field age M dwarf for developing stellar models that include chromospheres. \cite{Fontenla2016} and \cite{Peralta2023} found that using the revised non-LTE models and additional opacities would make the calculated spectra match the observed ones. Generally speaking, stellar photosphere models like PARSEC do match the broadband NUV fluxes of stars on the upper branch, but the inclusion of a model chromosphere is necessary to match prominent NUV emission lines. The fluxes of these chromospheric emission lines vary from star to star, requiring model chromospheres to be finely tuned to match observations.

\section{Flaring in the NUV}

\begin{figure}
    \centering
    \includegraphics[scale=0.5]{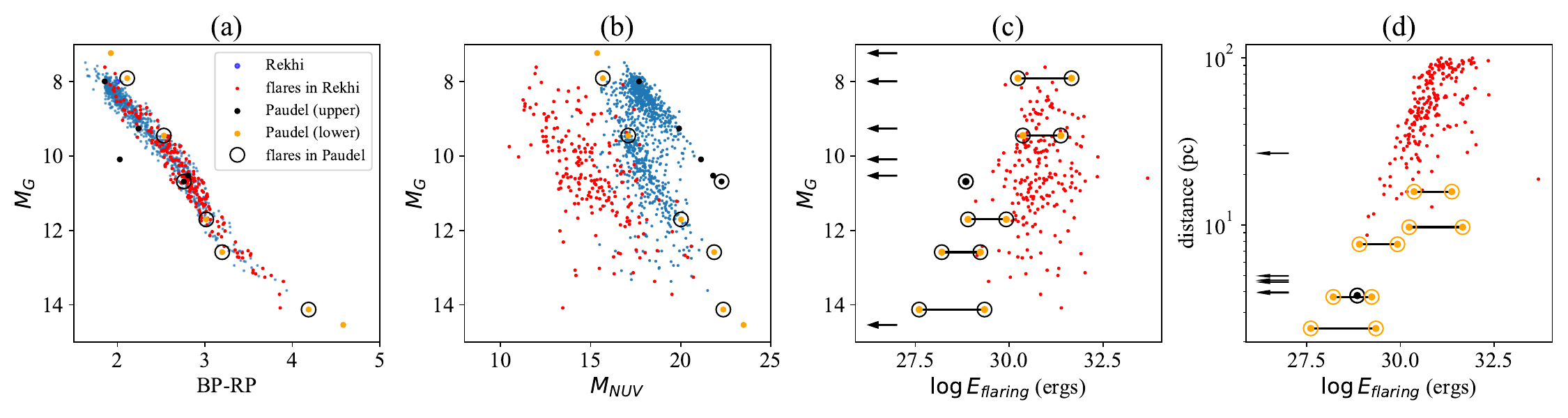}
    \caption{Flaring M dwarfs in the NUV band on various diagrams: the HRD (panel a), $M_{NUV}$ vs. $M_{G}$ (panel b), $\log$E$_{flaring}$ vs. $M_G{G}$ (panel c), and $\log$E vs. distance (panel d). Red and blue dots are stars with and without detectable NUV flaring in \cite{Rekhi2023}, respectively. Most of the NUV flaring stars in \cite{Rekhi2023} are on neither of the two branches. Orange and black dots represent stars in \cite{Paudel2024} on the lower and upper branches, respectively. Open circles highlight stars with detected NUV flares in \cite{Paudel2024}. Because multiple flares are reported in \cite{Paudel2024} for a given star, the largest and smallest flares are connected by a black line in panel c. Stars from \cite{Paudel2024} without detected NUV flares are shown in arrows marked at their $M_G$ magnitudes in panel c and at their distances in panel d.}
    \label{fig:GALEXflaring}
\end{figure}

M dwarf flaring has been discussed extensively in the optical band using data acquired from the {\it Kepler} or {\it TESS} missions \citep{Hawley2014, Davenport2016, Gunther2020}. Using the Kepler and K2 results reported in \cite{Liu2019}, \cite{Henry2024} found that $\sim$40\% of M dwarfs earlier than spectral type M5 show white-light flares. Recently, \cite{Rekhi2023} conducted a study of the NUV flaring of 4,176 M dwarfs on the main sequence selected from the {\it TESS} input catalog using the {\it GALEX} lightcurves. Among the 361 (8.6\%) NUV flaring M dwarfs in their sample, 207 (4.9\%) flaring stars meet the same selection criteria for the GAGDR3 sample, so the percentage of flaring stars appears to be much less than that of white-light flaring stars. Figure~\ref{fig:GALEXflaring} shows most of these NUV flaring M dwarfs are brighter in the NUV than stars on both branches. The range of flare energies ($E_{flaring}$) is reported between 1.3$\times10^{29}$ and 4.8$\times10^{33}$ erg in \cite{Rekhi2023}.


\cite{Paudel2024} also conducted a flaring study of selected 24 nearby M dwarfs observed by the {\it Swift} in the NUV, and only 12 stars have RUWE$<$1.4. Among these 12 stars, six flaring stars can be seen on both branches, but they are mostly fainter than those flaring stars in \cite{Rekhi2023} in the absolute NUV magnitude. The remaining six stars have no detected flares, so the percentage of NUV flaring stars is 50\%, similar to that of white-light flaring stars. \cite{Paudel2024} found the range of NUV flare energies for these nearby stars is between 4.0$\times10^{27}$ and 4.6$\times10^{31}$ erg. These two works are compared in Figure~\ref{fig:GALEXflaring}.

Apparently, the detected flaring energy ranges and percentage of flaring stars are different between these two works. We think the following two reasons could contribute to the differences we see here. First, targets in \cite{Paudel2024} are nearby M dwarfs, but most of the targets in \cite{Rekhi2023} are much further away. Therefore, small flares from stars afar won't be detected, as we can see in the panel d of Figure~\ref{fig:GALEXflaring}. Second, the NUV bandpasses and throughputs are different between {\it Swift} and {\it GALEX} given in Table~\ref{tbl:filters}.

\begin{deluxetable}{ccc}
\label{tbl:filters}
\tablecaption{NUV filters}
\tablehead{
\colhead{}&
\colhead{{\it Swift}/UVM2} &
\colhead{{\it GALEX}/NUV}
}
\colnumbers
\startdata
$\lambda_{min}$ (\AA) & 1699.1 & 1693.27  \\
$\lambda_{cen}$ (\AA) & 2259.8 & 2353.27  \\
$\lambda_{max}$ (\AA) & 2964.3 & 3007.56  \\
FWHM (\AA)            &  527.1 &  795.19  \\
the highest throughput & 2.9\% @ 2122.49\AA & 3.1\% at 2200\AA
\enddata
\end{deluxetable}

Finally, although the {\it Swift} has a smaller effective area than {\it GALEX}'s, targets observed by {\it Swift} have a much longer exposure time (40k seconds) than those observed by the {\it GALEX} All Sky Imaging Surveys mode (100 seconds). Consequently, 93\% of stars reported in \cite{Rekhi2023} have NUV flare energies greater than $10^{30}$ erg, but only 25\% of the NUV flare fluxes reported in \cite{Paudel2024} are greater than $10^{30}$. Apparently, only large flares tend to be detected by {\it GALEX}. For example, because stars in \cite{Rekhi2023} are further than 20 pc, some may be unresolved binaries, even though we only keep stars with low RUWE. Without knowing the binarity of each star, it is difficult to understand whether the detected large NUV flares are from the primary or unresolved secondary. Nonetheless, these stars appear to be extremely active in the NUV, and a potential third branch can be seen.

In Figure~\ref{fig:GALEXflaring}, five of the six stars with NUV flares in \cite{Paudel2024} are on the lower branch, and only one star on the upper branch, Luyten's star, has a detectable flare ($7\times10^{28}$ ergs). This shows even stars on the upper branch could still exhibit flares. Even though most stars in \cite{Paudel2024} don't have flaring detection, we expect more stars to have NUV flaring below the detection threshold of {\it GALEX} and {\it Swift}. Indeed, \cite{Paudel2024} use {\it HST}/STIS spectra to measure from GJ 832, an upper branch star, two small flares that would have been below the detection threshold of {\it Swift}. The bright detection limit from the {\it GALEX} could be the reason why the percentage of stars with NUV flares is much lower than it is in the optical band. Yet, systematically studying NUV activity for such a large all-sky sample is still impossible until the Ultra-violet Explorer (UVEX) launch in 2030 \citep{Kulkarni2021} with improved sensitivity over {\it GALEX}.

\section{Mg II doublet strength in the NUV band}
\label{sec:MgII}

The Mg II doublet at 2802.7 and 2795.5\AA~is one of the most prominent features in the NUV band and is often used to diagnose chromospheric activity. For F, G, and early K dwarfs, the Mg II emission line cores from the chromosphere are within broad absorption wings formed in the photosphere, and Mg II lines become emission without wings for stars cooler than early K dwarfs \citep{Linsky2017}. Recently, \cite{Pal2023} analyzed 513 stars observed using the Hubble Space Telescope's Space Telescope Imaging Spectrograph (STIS) CCD and G230LB grating in the NUV band, and their sample includes evolved giants and dwarfs from F to mid-M types. They measure the equivalent-width-style index of Mg II doublet by defining windows for the emission line and pseudo-continuum\footnote{This term of the ``equivalent-width-style'' index is defined in \cite{Pal2023}. However, although it is called an index, it has a unit of magnitude and differs from the Mount Wilson $S$ index in the study of the Ca II H and K lines, which also have emission lines within broad absorption wings.}. However, because the window they used to determine this index at the Mg II doublet is very wide (30\AA), stars hotter than early K dwarfs appear to have absorptions, and their core Mg II emission lines are embedded within those wings. Hence, their Mg II index is dominated by the absorption wings for stars hotter than early-K dwarfs. They found that the Mg II doublet index relates to B-V color, and the line absorption increases from A0 stars and peaks around G-type stars for both dwarfs and giants at $B-V\sim$0.65. For stars color than $B-V\sim$1.0 or K3 type, almost all stars have Mg II line emission as discussed in \cite{Linsky2017}. Therefore, the Mg II activity is related to the masses or spectral types.

\begin{figure}
    \centering
    \includegraphics[scale=0.57]{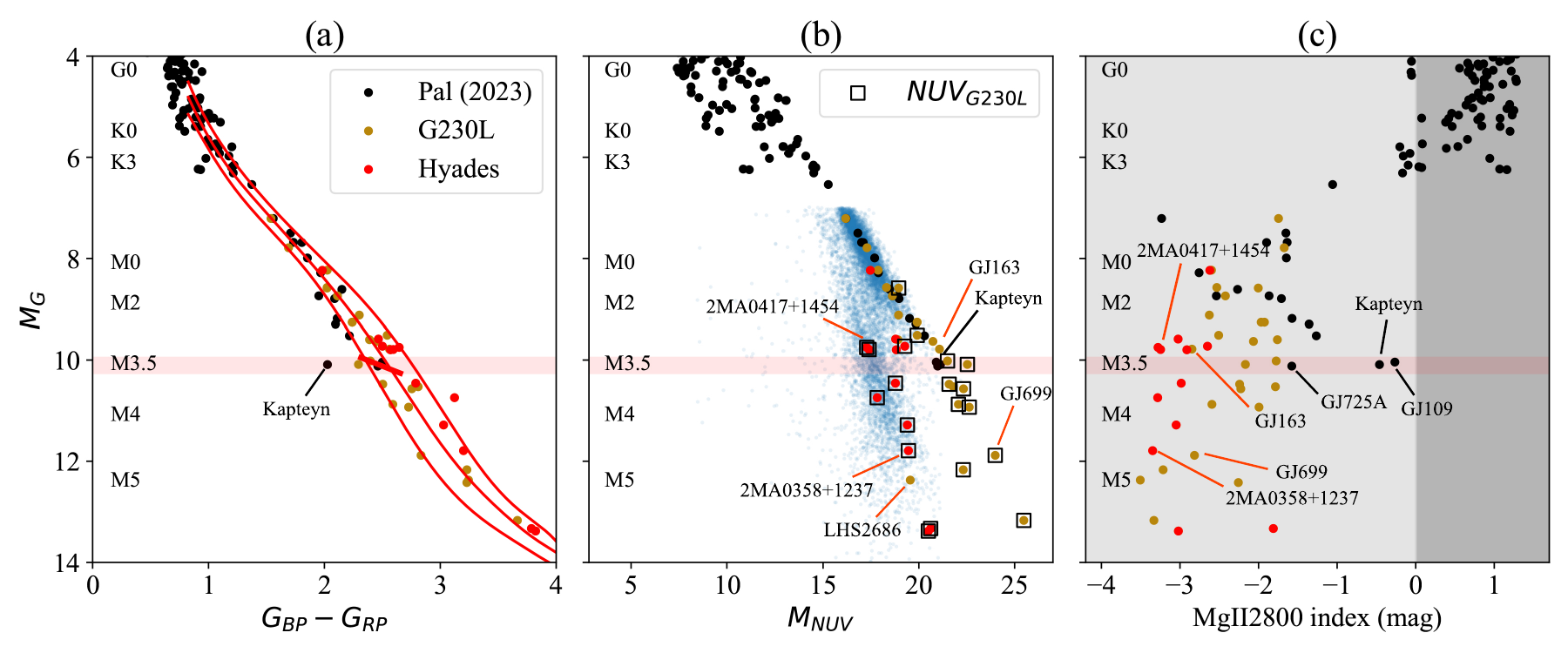}
    \caption{Stars with Mg II indices on HRD (panel a), $M_G$ vs. $M_{NUV}$ (panel b), and $M_G$ vs. Mg II index diagrams (panel c). Stars from \cite{Pal2023} are shown in black dots. An additional 38 stars retrieved from the MAST with the STIS-MAMA$+$G230L observing configuration are shown in brown dots. Red dots represent known members of the Hyades. Open boxes indicate stars without {\it GALEX} NUV photometry. In panel c, a dark-gray region indicates stars with positive indices of strong absorption wings and core Mg II emission, and a light-gray region indicates negative indices of strong Mg II emission from the chromosphere. Red lines in panel a are defined in Figure~\ref{fig:corner}, and blue dots in panel b are the GAGDR3 sample.}
    \label{fig:MgII}
\end{figure}

To make a consistent comparison with our analysis in this work, we match their stars with the {\it Gaia} DR3 and plot the Mg II index against $M_G$ shown in Figure~\ref{fig:MgII} instead of $B-V$ color used in \cite{Pal2023}. Besides, to supplement additional cool M dwarfs in this figure, we retrieve M dwarfs observed using the STIS-MAMA and G230L configurations with 1-D flux calibrated spectra from MAST using revised automatic coadding algorithm \citep{Debes2024}. We then follow the algorithm defined in \cite{Pal2023} to calculate the Mg II index for these additional 38 stars given in Table~\ref{tbl:MgII}. Some of these stars don't have NUV photometry in {\it GALEX} shown as open boxes in Figure~\ref{fig:MgII}, so their NUV magnitudes are calculated by convolving the STIS spectra with the {\it GALEX} NUV bandpass. The equation to calculate {\it GALEX} NUV photometry and the NUV bandpass are available at the \href{https://asd.gsfc.nasa.gov/archive/galex/FAQ/counts_background.html}{{\it GALEX} FAQ page} and the \href{http://svo2.cab.inta-csic.es/svo/theory/fps3/index.php?id=GALEX/GALEX.NUV}{Spanish Virtual Observatory (SVO) filter service}.

We found that stars hotter than $\sim$K3 dwarfs mostly show Mg II absorption, but all stars cooler than these early K dwarfs have emission. Interestingly, although the Mg II line strength increases by 4 magnitudes from 1 of K0 to $-$3 of early M, the absolute NUV magnitudes continue to decrease and follow the trend line for those stars on the upper branch in Figure~\ref{fig:analysis}. This indicates the additional excess flux must come from other sources to move stars to the lower branch.

We highlight three mid-M dwarfs, GJ 109, GJ 725A, and Kapteyn's star, in Figure~\ref{fig:MgII}. These three stars have similar $M_G$ and $M_{NUV}$ on the upper branch, clustering at the location of the Kapteyn's star in the $M_G$ vs. $M_{NUV}$ plot in Figure~\ref{fig:MgII}(b). Because of the crowdedness of this plot, we only label Kapteyn's star. Two of the three stars, GJ 109 and GJ 725A, have similar metallicities, but Kapteyn's star is a nearby metal-poor subdwarf. Hence, Kapteyn's star is off to the left side of the main sequence in panel a, and the other two stars are close to the main-sequence gap. Table~\ref{tbl:threestars} summarizes their parameters. 

Although they have similar $M_{NUV}$ and are inactive at the H$\alpha$, all three stars have Mg II emission, and their Mg II indices differ by as much as 1.3 mags. Their rotation periods seem not to follow the trend of the Mg II index, where the fastest rotator is GJ 109, but GJ 109 has the weakest Mg II line. Kapteyn's star, a slow rotator, even has a stronger Mg II line than GJ 109 has. Based on this very limited sample, it appears that the strength of the Mg II line won't be a major contributor to making stars bright in the NUV. In the next section, we will discuss the possible source of the excess NUV flux at a given $M_G$.

\begin{deluxetable}{ccccc}
\label{tbl:threestars}
\tablecaption{Highlighted three mid-M dwarfs}
\tablehead{
\colhead{}&
\colhead{GJ109} &
\colhead{GJ725A} &
\colhead{Kapteyn's star} &
\colhead{Ref}
}
\colnumbers
\startdata
$M_G$              & 10.04  &  10.12   & 10.09 & \\
$M_{NUV}$          & 20.91  & 20.98    & 21.13 & \\
$[Fe/H]$             & $-$0.2   &  $-$0.31   &  $-$0.99 & 2, 3 \\
Mg II index (mags) & $-$0.265 &  $-$1.577  &  $-$0.461 & 1\\
rotation (days) & 38.9   &  103.5  &  143 & 4, 5, 6 \\
H$\alpha$   &  inactive   &   inactive   & inactive & 4, 7
\enddata
\tablecomments{This table is ordered by rotation periods from left to right. 1: \cite{Pal2023}, 2: \cite{Bonfils2005}, 3: \cite{Woolf2005}, 4: \cite{Schofer2019}, 5: \cite{Fouque2023}, 6: \cite{Newton2016}, 7: \cite{Kirkpatrick1991}}
\end{deluxetable}

\begin{deluxetable}{ccccccccccccc}
\tabletypesize{\tiny}
\label{tbl:MgII}
\tablecaption{M dwarf Mg II indices from the MAST}
\tablehead{
\colhead{Data} & 
\colhead{Name} & 
\colhead{RA} &
\colhead{Decl.} & 
\colhead{parallax} & 
\colhead{pmRA} &
\colhead{pmDE} & 
\colhead{RUWE} &
\colhead{Gmag} & 
\colhead{BP-RP} & 
\colhead{Mg II index} &
\colhead{$M_{NUV}$} &
\colhead{in GALEX}\\
\colhead{} & 
\colhead{} & 
\colhead{(deg)} &
\colhead{(deg)} & 
\colhead{(mas)} & 
\colhead{(mas)} &
\colhead{(mas)} & 
\colhead{} &
\colhead{(mag)} & 
\colhead{(mag)} & 
\colhead{} &
\colhead{(mag)} &
\colhead{}
}
\colnumbers
\startdata 
ODO306010&2MASSJ02564122+3522346&44.173&35.37599&26.066&200.986&-63.298&1.075&12.5&-3.02&-3.02&18.79&1\\
ODO307010&2MASSJ03553690+2118482&58.90488&21.31313&34.7283&206.778&-60.415&1.314&15.68&-3.02&-3.02&20.50&0\\
ODO308010&2MASSJ03555715+1825564&58.98887&18.43215&22.8552&136.739&-30.792&1.277&13.0&-2.91&-2.91&18.82&1\\
ODO309010&2MASSJ03581434+1237408&59.56044&12.62797&21.9478&129.707&-12.256&1.225&15.08&-3.35&-3.35&19.46&0\\
ODO366010&2MASSJ04110642+1247481&62.7774&12.79666&23.0698&132.354&-13.39&0.782&16.51&-1.81&-1.81&20.62&0\\
ODO305010&2MASSJ04172811+1454038&64.36774&14.90098&20.283&105.84&-17.354&1.235&13.26&-3.25&-3.25&17.40&0\\
ODO312010&2MASSJ04183382+1821529&64.64158&18.36458&21.821&116.932&-31.226&1.197&14.59&3.03&-3.05&19.40&0\\
ODO304010&2MASSJ04202761+1853499&65.11561&18.89706&19.673&104.719&-32.57&1.279&13.99&-2.98&-2.98&18.78&0\\
ODO317010&2MASSJ04254182+1900477&66.4247&19.0131&16.1683&84.101&-25.852&1.085&13.68&-2.65&-2.65&19.26&0\\
ODO313020&2MASSJ04262170+1800009&66.5909&18.00015&19.4275&97.858&-29.734&1.151&14.3&-3.28&-3.28&17.83&0\\
ODO314010&2MASSJ04290015+1620467&67.25112&16.34621&21.6325&105.107&-27.317&1.221&13.07&-3.28&-3.28&17.28&0\\
ODO351010&2MASSJ04483062+1623187&72.12807&16.3884&20.7646&86.517&-30.319&1.126&11.65&-2.62&-2.62&17.47&1\\
OEOO10020&BD-10-47&4.77301&-9.96619&49.4757&-36.325&-301.491&0.908&9.32&-1.67&-1.67&17.29&1\\
OEOO01030&BD-17-588A&45.46242&-16.59453&145.6922&-369.972&-267.931&1.073&10.06&-2.59&-2.59&22.07&0\\
OEOO17030&CD-23-12010&225.07985&-24.45421&36.3525&-199.483&-27.997&1.092&9.41&-1.75&-1.75&16.18&1\\
OEOO13040&CD-60-8051&350.0285&-60.06575&39.7113&-319.924&-127.782&1.179&10.24&-2.6&-2.6&17.87&1\\
OED610010&DENIS-J104814.6-395606&162.0539&-39.93963&247.2156&-1179.311&-988.121&1.045&14.01&-2.73&-2.73&25.87&0\\
ODSX03030&GJ-1061&54.00339&-44.51436&272.1615&745.654&-373.323&1.192&11.0&-3.33&-3.33&25.48&0\\
OBPI10010&GJ1214&258.83148&4.96058&68.2986&580.202&-749.713&1.009&13.0&-3.21&-3.21&22.32&0\\
ODLM15020&GJ163&62.32308&-53.3711&66.0705&1046.236&584.166&1.371&10.68&-2.85&-2.85&21.08&1\\
OCK125020&GJ176&70.73548&18.9532&105.4275&656.647&-1116.594&0.888&9.0&-2.63&-2.63&18.94&1\\
OCK128020&GJ436&175.55067&26.70295&102.3014&895.088&-813.55&1.125&9.58&-2.07&-2.07&20.73&1\\
ODLM12020&GJ649&254.5363&25.7419&96.2333&-115.314&-508.087&1.164&8.82&-2.42&-2.42&18.62&1\\
OCK122020&GJ667C&259.75125&-34.9978&138.0663&1131.517&-215.569&0.956&9.39&-2.17&-2.17&22.53&0\\
ODLM24020&GJ699&269.4485&4.73942&546.9759&-801.551&10362.394&1.085&8.19&-2.82&-2.82&23.99&0\\
OCK116010&GJ832&323.39124&-49.01263&201.325198&-45.917&-816.875&1.097&7.74&-1.97&-1.97&19.91&1\\
ODLM18020&GJ849&332.42316&-4.64083&113.4447&1132.583&-22.157&1.354&9.24&-2.51&-2.51&19.91&0\\
OBPI04020&GJ876&343.32411&-14.26669&214.038&957.715&-673.601&1.340&8.88&2.81&-1.79&21.80&1\\
ODSX01010&HD-204961&323.39125&-49.01263&201.325&-45.917&-816.875&1.097&7.74&2.24&-1.93&19.91&1\\
OEOO05020&HD-304636&140.39914&-60.28114&95.6982&-840.185&182.056&1.021&8.68&-2.0&-2.0&18.94&0\\
OEOO09010&L-22-69&290.23831&-82.55996&80.113&342.300&-1230.297&0.961&11.41&2.73&-1.99&22.63&0\\
OECB05050&L-98-59&124.53290&-68.31451&94.266&94.794&-340.084&1.272&10.61&2.50&-2.24&21.58&0\\
OECB20020&L-678-39&144.00749&-21.6652&105.9789&138.722&-990.342&1.119&9.89&-1.78&-1.78&21.50&0\\
ODLM34020&LHS-2686&197.54842&47.75245&82.039&-636.059&-616.392&1.390&12.80&3.24&-3.51&19.55&1\\
ODLM30020&LP-756-18&313.91116&-14.06734&80.2048&1420.755&-472.266&1.262&12.9&-2.26&-2.26&...&0\\
OEOO15030&LP-961-53&178.57803&-37.55338&36.8291&250.996&-144.947&1.138&10.74&-2.53&-2.53&18.31&1\\
OEOO02030&WOLF-437&191.98138&9.74935&123.7756&-1008.267&-460.034&1.094&10.11&-2.23&-2.23&22.33&0\\
OEDF05010&UCAC3-49-21611&97.09568&-65.57859&32.133&-102.641&161.748&1.353&12.06&2.39&-1.76&18.91&1
\enddata
\tablecomments{This table is sorted by the target name. The first column is the unique observation identifier (``obsID")  associated with the STIS observation in the MAST. The last column indicates the source of the NUV magnitudes: ``1'' is from {\it GALEX} and ``0'' is calculated from the STIS spectra. LP 756-18 has a truncated STIS spectrum below the 1750\AA, so the NUV magnitude can't be calculated. This table is ordered alphabetically by the target names.}
\end{deluxetable}

\section{NUV spectra between branches}
\label{sec:HST_NUV}

To discern the distinctions between the two branches, we selected two pairs of stars in panel b of Figure~\ref{fig:MgII} with similar luminosities or $M_G$ magnitudes but located on different branches. Their {\it HST}/STIS spectra are retrieved from the MAST. This shows how NUV spectra change at a given luminosity, assuming comparable masses. The first pair consists of GJ 163 ($M_G=$9.78, $M_{NUV}=$21.07) and 2MASS J04172811$+$1454038 (hereafter 2MA0417$+$1454, $M_G=$9.79, $M_{NUV}=$17.47). The second pair comprises GJ 699 (Barnard's star, $M_G$=11.88, $M_{NUV}=$23.99) and 2MASS J03581434$+$1237408 (hereafter 2MA0358$+$1237, $M_G$=11.79, $M_{NUV}=$19.46). Both 2MASS stars are members of Hyades, and their NUV spectra are presented in Figure~\ref{fig:compare}.

In Figure~\ref{fig:compare}, we can see that two 2MASS stars exhibit additional flux from the Fe II emission line forest at $\sim$2400 and 2600\AA~and a stronger Mg II doublet. Because the Mg II doublet lies in the low throughput region of the {\it GALEX} NUV filter, the accumulated excess flux, integrated from the shorter to longer wavelengths, primarily originates from the Fe II lines, as shown in the bottom two plots of Figure~\ref{fig:compare}. Unlike the first pair, the $\Delta M_G$ of the second pair is larger, suggesting they may not have the same masses. Nevertheless, both 2MASS stars still exhibit higher fluxes primarily from the Fe II lines than their counterparts. As for the Mg II doublet, although it is strong in both 2MASS stars, it contributes less than 20\% of accumulated excess fluxes. Therefore, this result is consistent with the discussion about Mg II indices of the three mid-M dwarfs in section~\ref{sec:MgII} that the Mg II doublet is not a major contributor to the excess flux.

Both GJ 163 and GJ 699 are slow rotators, with rotational periods of 87 and 130.4 days, respectively \citep{Astudillo2017, Newton2017}. Conversely, due to their youth, 2MA0417$+$1454 and 2MA0358$+$1237 are fast rotators, with rotational periods of 2.38 and 0.87 days, respectively \citep{Douglas2016}. Fast rotation may be partly responsible for the enhanced brightness of stars in the NUV band, as illustrated in Figure~\ref{fig:rotation}. 

\cite{Kowalski2019} compared NUV fluxes of GJ 1243, an active nearby flaring M dwarf, between quiescent and flaring states. They found a brightening of NUV fluxes during flares is from both emission lines and continuum. However, flux differences between the two branches seen in Figure~\ref{fig:compare} mainly manifest at emission lines, with the continuum difference remaining close to zero. Thus, the brightening mechanisms between quiescent-to-flaring states and lower-to-upper branches likely differ.

\section{Conclusion}

M dwarfs host an abundance of exoplanets, and their NUV light drives photochemistry \citep{Tarter2007, Rugheimer2015} and may be harmful or helpful to surface life \citep{Rugheimer2015b, Rimmer2018}. Thus, accurately characterizing their activities in the NUV band is important to understand ``Worlds and Suns in Context,'' a key priority area identified by the National Academies’ ``Pathways to Discovery in Astronomy and Astrophysics for the 2020s'' decadal survey \citep{Astro2020}. In this study, we report two populations of M dwarfs in the NUV band using the HLSP catalog of cross-matched {\it GALEX} and {\it Gaia} sources from the MAST \citep{Bianchi2020}. The two populations are mainly divided by their levels of NUV flux as seen on the $M_{NUV}$ vs. $M_G$ diagram in Figure~\ref{fig:corner}.  The ``upper'' branch refers to low NUV fluxes, and the ``lower'' branch refers to higher NUV fluxes. The increase in the number of stars on the lower branch starts approximately around the spectral type M2, coinciding with a $NUV-G_{RP}$ color shift previously thought to be caused by the M dwarf's interior transition \citep{Cifuentes2020}. In order to understand the anomaly at the M2 type and these two branches, we initiate various studies based on results from literature and archival data.

The mass-luminosity relation has long been known to exhibit a kink near M2, explained by the formation of H2 and increased efficiency in convective energy transport. However, it is unclear whether that is related to the NUV anomaly shown in Figure~\ref{fig:corner}. In section~\ref{sec:HST_NUV}, we show that the excess NUV flux between the upper and lower branches is mainly from the Fe II lines, and the continuum remains the same. Consequently, the connection between $H_2$ formation and excess NUV flux at M2 or strong Fe II lines is yet to be studied. We also find that the excess fluxes or gap between the two branches increase as masses decrease, reaching factors of 3$\times$ to 25$\times$ of the total fluxes in the NUV for stars on the upper branch. The difference grows much larger in the FUV band. We find youth and fast rotation may stimulate the NUV activity for stars on the lower branch. However, as we show in Figure~\ref{fig:HAZMAT},~\ref{fig:Halpha}, and \ref{fig:rotation}, not all stars on the lower branch are young, exhibit active H$\alpha$ emission or are fast rotators. We don't know the exact cause of these excess NUV fluxes for these relatively inactive stars, but unresolved binaries could be the main reason. Other scenarios, such as non-linear propagation of Alfvén waves from the photosphere to the upper atmosphere \citep{Sakaue2021}, or star-planet interaction \citep{Saur2013} may also induce the upper atmosphere heating to contribute some degrees of excess fluxes.

Studying M dwarf flares in the NUV is limited by available UV telescopes, their sensitivities, and observing strategies. Using results from \cite{Rekhi2023} and \cite{Paudel2024}, we find the percentage of M dwarfs with flares in the NUV ranges from a few percent to 50\%, depending on the telescope sensitivity and total stare time per target. We also find stars showing NUV flares acquired by the {\it GALEX} are predominantly brighter than stars on both branches in the NUV, suggesting a possible third branch of stars. On the other hand, stars with NUV flares detected by the {\it Swift} can be found on both branches. The different instrument designs and survey strategies between these two telescopes could be the cause of different flaring energies being detected. Based on results from \citet{Paudel2024} using {\it Swift}, one star on the upper branch flares in the NUV, so we expect more stars on the upper branch to flare, but no existing dataset or telescope can systematically study NUV flares for a large number of M dwarfs on both branches.

The two branches of M dwarfs identified photometrically in the NUV can also be studied spectroscopically to uncover what emission lines and continuum contribute to the variations in NUV photometry. Given that HST serves essentially as the major source of NUV spectroscopic observations of M dwarfs available to us, Figure~\ref{fig:MgII} summarizes the current status of the archival NUV spectra using specific HST instrument configurations (STIS/NUV-MAMA$+$G230L and STIS/NUV-MAMA$+$G230LB, hereafter LLB). The LLB configuration can produce a full spectral coverage in the NUV. In contrast, HST NUV spectra obtained with the Cosmic Origins Spectrograph (COS) contain a spectral gap between 1750 and 2800\AA. For example, \cite{Peacock2020} used HST/COS to study the evolution of UV spectra of early M dwarfs with estimated ages between 10 and 5000 Myrs. They generate synthetic spectra to match major UV lines like Ly$\alpha$, Mg II, and pseudo-continuum. However, due to the spectral gap in the COS data, only a small fraction of the spectral range in the NUV band is available to compare with their models. 

\begin{figure}
    \centering
    \includegraphics[scale=0.4]{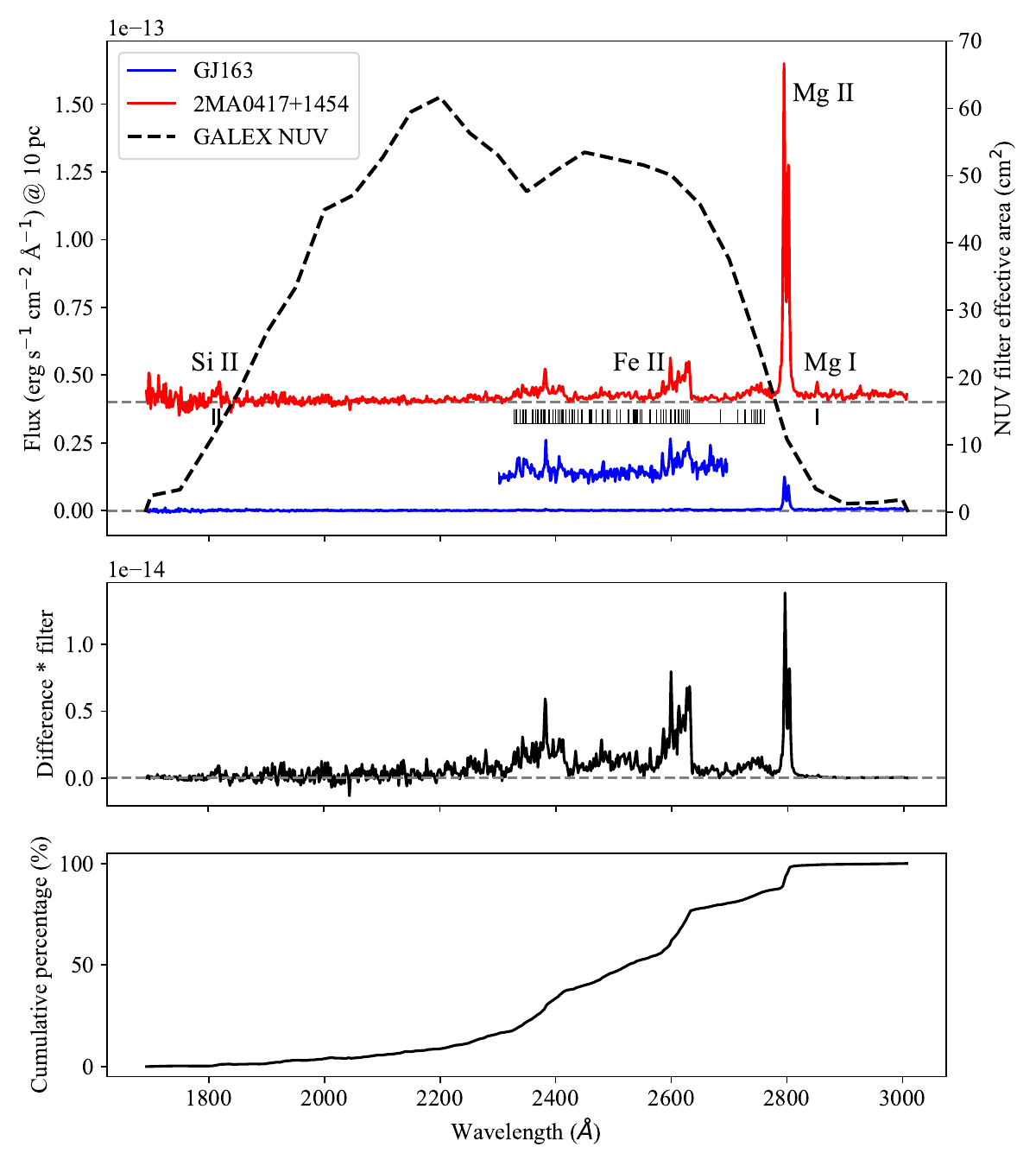}    
    \includegraphics[scale=0.4]{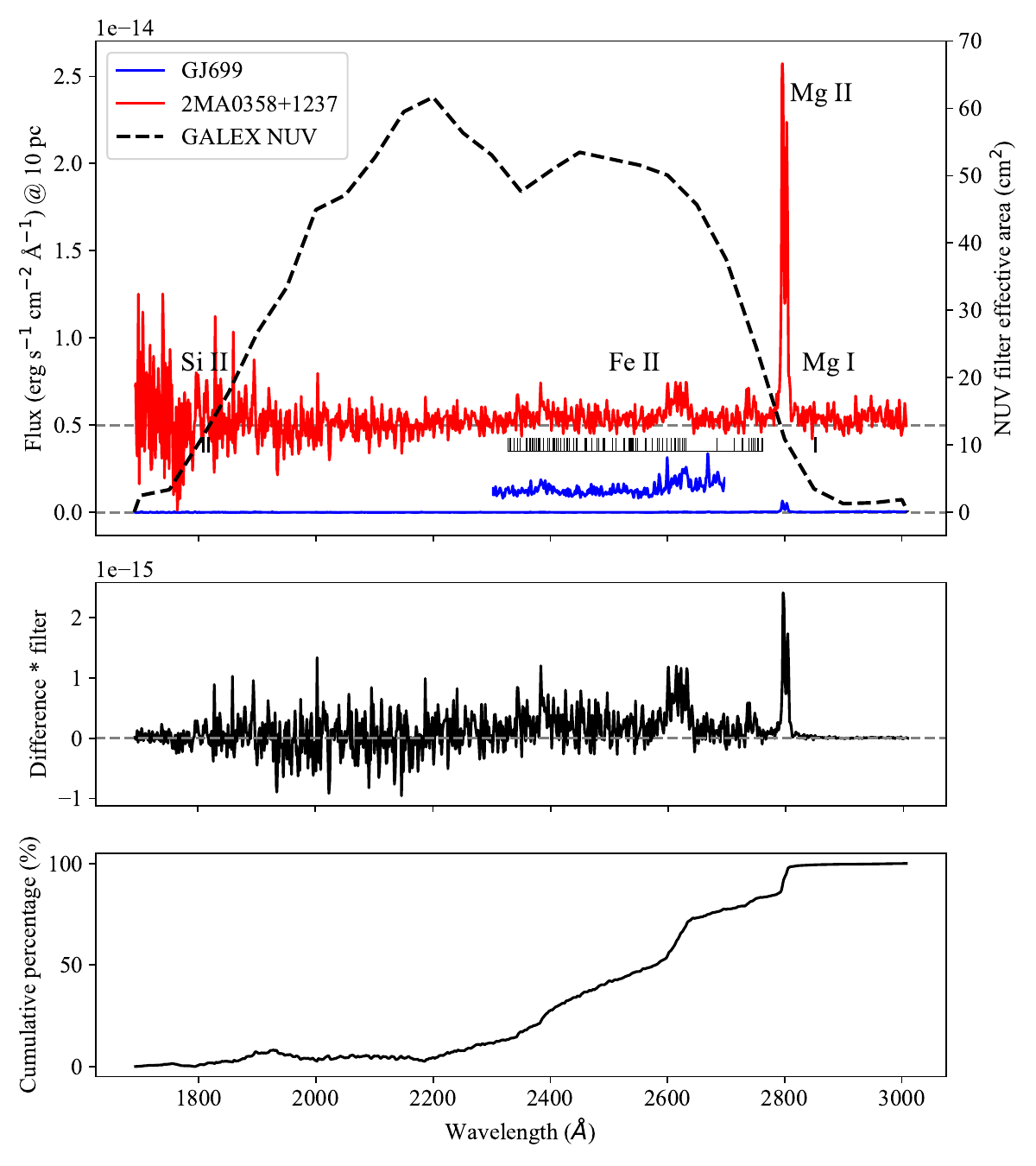}
    \caption{Spectral comparisons between stars on different branches. The left column compares spectra for GJ163 and 2MA0417+1454, and the right column is for GJ699 and 2MA0358+1237. (Top) Spectra are calibrated to a distance of 10 pc, with the red spectra shifted upward relative to the blue spectra for better visual comparison. The short blue spectra magnify the Fe II forests from the fainter comparison stars. Key features such as Si II, Fe II, and Mg I lines are marked with ticks. The strong emission at 2800\AA~is the Mg II doublets. The black dashed lines indicate the {\it GALEX} NUV bandpass, with its filter effective area labeled on the right side of the top two plots. (Middle) Spectral differences for each pair, convolved with the {\it GALEX} NUV bandpass. (Bottom) Cumulative percentage of flux from the middle plots.}
    \label{fig:compare}
\end{figure}

We study the Mg II line strength from early G to late M dwarfs using both the HST archival data with LLB configuration and results in \cite{Pal2023}. We find essentially all dwarfs cooler than $\sim$K3 type exhibit Mg II emission (Figure~\ref{fig:MgII}). Therefore, we conclude that all M dwarfs on both branches must have Mg II emission. To understand the flux gap between the two branches (excess flux), we select two pairs of stars with similar $M_G$, one from the upper branch and one from the lower branch, shown in Figure~\ref{fig:compare}. Our examination suggests that these excess fluxes are dominated by the Fe II lines in the NUV, with fast rotators potentially exhibiting stronger lines, thus enhancing the brightness in the NUV band. On the contrary, the strong Mg II line lies near the edge of the {\it GALEX} NUV bandpass and only contributes less than 20\% of excess fluxes between the two branches.

The upper branch has been more thoroughly characterized spectroscopically than the lower branch, limiting the comparison between the two. In particular, lower branch stars with LLB spectra are almost exclusively mid-type M dwarfs and Hyades members, shown in Figure~\ref{fig:MgII}.  We think such a selection-biased sample could be caused by the M dwarf bright objection restrictions put in place to protect the safety of HST's UV detectors (Instrument Science Report STIS 2017-02 by Osten).  This biased sample limits the following two spectral analyses: 1) a spectral comparison between stars on the two branches at a given $M_G$ or mass, and 2) establishing spectral sequences in the NUV band for stars on the lower and upper branches. 

The first analysis, similar to spectral comparisons given in Figure~\ref{fig:compare}, could help us to understand what features contribute to the excess fluxes at a given mass. However, most of the stars on the lower branch in the HST archive are from the Hyades, so there is no diverse sample. Figure~\ref{fig:HAZMAT} demonstrates that many stars on the lower branch, while not associated with any young moving groups, still exhibit excess NUV fluxes. If these stars meet the HST NUV bright object restrictions, they would be ideal targets for the HST to observe in the future. Furthermore, we lack early M dwarfs on the lower branch with a full spectral coverage in the NUV band for comparative analysis across the two branches. 

The second spectral analysis resembles establishing a spectral sequence in the optical band, such as TiO band or H$\alpha$ line against effective temperature \citep{Hawley1996, West2011, Schmidt2015}, so that we can understand relations between various lines' features, temperatures, and ages. For example, \cite{Linsky2020} showed that the ratios of X-ray over Ly$\alpha$ luminosities and the stellar bolometric luminosities increase as effective temperatures decrease. However, that study is limited to Ly$\alpha$, which is beyond the {\it GALEX} NUV coverage. Similar studies could be implemented on stars on the upper branch because those stars are well covered from high to low masses in the HST archival data. Yet, this kind of study is currently limited by a few stars on the lower branch in the HST archive. Therefore, increasing the number of stars with a full spectral coverage in the NUV band on the lower branch is essential and could assist us in understanding not only the spectral sequence but also the underlying physics of the increasing number of M dwarfs on the lower branch at spectral type M2.

\bigskip
We thank the anonymous referee for constructive comments that improved the original manuscript. We also thank Sarah Peacock, Kevin France, and Greg Feiden for their suggestions. This research has made use of NASA’s Astrophysics Data System. The GAGDR3 and HST NUV spectra presented in this paper were obtained from the Multimission Archive at the Space Telescope Science Institute (MAST). STScI is operated by the Association of Universities for Research in Astronomy, Inc., under NASA contract NAS5-26555. Support for MAST for non-HST data is provided by the NASA Office of Space Science via grant NAG5-7584 and by other grants and contracts. This research has made use of the "Aladin sky atlas", SIMBAD database, and VizieR catalogue, operated at CDS, Strasbourg, France. This research has made use of the Spanish Virtual Observatory (https://svo.cab.inta-csic.es) project funded by MCIN/AEI/10.13039/501100011033/ through grant PID2020-112949GB-I00. This research has made use of the Spanish Virtual Observatory (https://svo.cab.inta-csic.es) project funded by MCIN/AEI/10.13039/501100011033/ through grant PID2020-112949GB-I00. This work has made use of data from the European Space Agency (ESA) mission {\it Gaia} (\url{https://www.cosmos.esa.int/gaia}), processed by the {\it Gaia} Data Processing and Analysis Consortium (DPAC, \url{https://www.cosmos.esa.int/web/gaia/dpac/consortium}).  Funding for the DPAC has been provided by national institutions, in particular the institutions participating in the {\it Gaia} Multilateral Agreement.

%



\software{Matplotlib
  \citep{Hunter2007},
  NumPy \citep{vanderWalt2011}, astropy \citep{Astropy2013, Astropy2018},  SciPy \citep{Virtanen2020}, TOPCAT \citep{TOPCAT}, and scikit-learn \citep{scikit}}






\begin{thebibliography}{}

\bibitem[Astudillo-Defru et al.(2017)]{Astudillo2017} Astudillo-Defru, N., Delfosse, X., Bonfils, X., et al.\ 2017, \aap, 600, A13. doi:10.1051/0004-6361/201527078

\bibitem[Astropy Collaboration et al.(2013)]{Astropy2013} Astropy Collaboration, Robitaille, T.~P., Tollerud, E.~J., et al.\ 2013, \aap, 558, A33. doi:10.1051/0004-6361/201322068

\bibitem[Astropy Collaboration et al.(2018)]{Astropy2018} Astropy Collaboration, Price-Whelan, A.~M., Sip{\H{o}}cz, B.~M., et al.\ 2018, \aj, 156, 123. doi:10.3847/1538-3881/aabc4f

\bibitem[Basri(2022)]{Basri2022} Basri, G.\ 2022, The 21st Cambridge Workshop on Cool Stars, Stellar Systems, and the Sun, 114. doi:10.5281/zenodo.7530476

\bibitem[Benedict et al.(2016)]{Benedict2016} Benedict, G.~F., Henry, T.~J., Franz, O.~G., et al.\ 2016, \aj, 152, 141. doi:10.3847/0004-6256/152/5/141

\bibitem[Berger et al.(2020)]{Berger2020} Berger, T.~A., Huber, D., van Saders, J.~L., et al.\ 2020, \aj, 159, 280. doi:10.3847/1538-3881/159/6/280

\bibitem[Bianchi et al.(2017)]{Bianchi2017} Bianchi, L., Shiao, B., \& Thilker, D.\ 2017, \apjs, 230, 24. doi:10.3847/1538-4365/aa7053

\bibitem[Bianchi \& Shiao(2020)]{Bianchi2020} Bianchi, L. \& Shiao, B.\ 2020, \apjs, 250, 36. doi:10.3847/1538-4365/aba2d7


\bibitem[Bonfils et al.(2005)]{Bonfils2005} Bonfils, X., Delfosse, X., Udry, S., et al.\ 2005, \aap, 442, 635. doi:10.1051/0004-6361:20053046

\bibitem[Boyajian et al.(2012)]{Boyajian2012} Boyajian, T.~S., von Braun, K., van Belle, G., et al.\ 2012, \apj, 757, 112. doi:10.1088/0004-637X/757/2/112



\bibitem[Casagrande et al.(2008)]{Casagrande2008} Casagrande, L., Flynn, C., \& Bessell, M.\ 2008, \mnras, 389, 585. doi:10.1111/j.1365-2966.2008.13573.x


\bibitem[Chabrier \& Baraffe(2000)]{Chabrier2000} Chabrier, G. \& Baraffe, I.\ 2000, \araa, 38, 337. doi:10.1146/annurev.astro.38.1.337

\bibitem[Chen et al.(2014)]{Chen2014} Chen, Y., Girardi, L., Bressan, A., et al.\ 2014, \mnras, 444, 2525. doi:10.1093/mnras/stu1605

\bibitem[Chevalier et al.(2023)]{Chevalier2023} Chevalier, S., Babusiaux, C., Merle, T., et al.\ 2023, \aap, 678, A19. doi:10.1051/0004-6361/202347111


\bibitem[Cifuentes et al.(2020)]{Cifuentes2020} Cifuentes, C., Caballero, J.~A., Cort{\'e}s-Contreras, M., et al.\ 2020, \aap, 642, A115. doi:10.1051/0004-6361/202038295

\bibitem[Copeland et al.(1970)]{Copeland1970} Copeland, H., Jensen, J.~O., \& Jorgensen, H.~E.\ 1970, \aap, 5, 12

\bibitem[Davenport(2016)]{Davenport2016} Davenport, J.~R.~A.\ 2016, \apj, 829, 23. doi:10.3847/0004-637X/829/1/23

\bibitem[Debes et al.(2024)]{Debes2024} Debes, J., Sankrit, R., Fischer, T., et al.\ 2024, Instrument Science Report COS 2024-01, 31 pages




\bibitem[Douglas et al.(2016)]{Douglas2016} Douglas, S.~T., Ag{\"u}eros, M.~A., Covey, K.~R., et al.\ 2016, \apj, 822, 47. doi:10.3847/0004-637X/822/1/47



\bibitem[Fontenla et al.(2016)]{Fontenla2016} Fontenla, J.~M., Linsky, J.~L., Garrison, J., et al.\ 2016, \apj, 830, 154. doi:10.3847/0004-637X/830/2/154

\bibitem[Fouqu{\'e} et al.(2023)]{Fouque2023} Fouqu{\'e}, P., Martioli, E., Donati, J.-F., et al.\ 2023, \aap, 672, A52. doi:10.1051/0004-6361/202345839

\bibitem[France et al.(2016)]{France2016} France, K., Loyd, R.~O.~P., Youngblood, A., et al.\ 2016, \apj, 820, 89. doi:10.3847/0004-637X/820/2/89

\bibitem[Froning et al.(2019)]{Froning2019} Froning, C.~S., Kowalski, A., France, K., et al.\ 2019, \apjl, 871, L26. doi:10.3847/2041-8213/aaffcd

\bibitem[Gagn{\'e} et al.(2018)]{Gagne2018} Gagn{\'e}, J., Mamajek, E.~E., Malo, L., et al.\ 2018, \apj, 856, 23. doi:10.3847/1538-4357/aaae09

\bibitem[Gagn{\'e} \& Faherty(2018)]{Gagne2018b} Gagn{\'e}, J. \& Faherty, J.~K.\ 2018, \apj, 862, 138. doi:10.3847/1538-4357/aaca2e

\bibitem[Gaia Collaboration et al.(2021)]{GaiaDR3} Gaia Collaboration, Brown, A.~G.~A., Vallenari, A., et al.\ 2021, \aap, 649, A1. doi:10.1051/0004-6361/202039657

\bibitem[G{\"u}nther et al.(2020)]{Gunther2020} G{\"u}nther, M.~N., Zhan, Z., Seager, S., et al.\ 2020, \aj, 159, 60. doi:10.3847/1538-3881/ab5d3a

\bibitem[Hawley et al.(1996)]{Hawley1996} Hawley, S.~L., Gizis, J.~E., \& Reid, I.~N.\ 1996, \aj, 112, 2799. doi:10.1086/118222

\bibitem[Hawley et al.(2014)]{Hawley2014} Hawley, S.~L., Davenport, J.~R.~A., Kowalski, A.~F., et al.\ 2014, \apj, 797, 121. doi:10.1088/0004-637X/797/2/121

\bibitem[Henry \& Jao(2024)]{Henry2024} Henry, T.~J. \& Jao, W.-C.\ 2024, \araa, 62, 593. doi:10.1146/annurev-astro-052722-102740

\bibitem[Hunter (2007)]{Hunter2007} Hunter, J.D., \ 2007, Computing in Science \& Engineering, 9, 90


\bibitem[Jao et al.(2018)]{Jao2018} Jao, W.-C., Henry, T.~J., Gies, D.~R., et al.\ 2018, \apjl, 861, L11. doi:10.3847/2041-8213/aacdf6

\bibitem[Jao et al.(2023)]{Jao2023} Jao, W.-C., Henry, T.~J., White, R.~J., et al.\ 2023, \aj, 166, 63. doi:10.3847/1538-3881/ace2bb

\bibitem[Jeffers et al.(2018)]{Jeffers2018} Jeffers, S.~V., Sch{\"o}fer, P., Lamert, A., et al.\ 2018, \aap, 614, A76. doi:10.1051/0004-6361/201629599



\bibitem[Kirkpatrick et al.(1991)]{Kirkpatrick1991} Kirkpatrick, J.~D., Henry, T.~J., \& McCarthy, D.~W.\ 1991, \apjs, 77, 417. doi:10.1086/191611

\bibitem[Kowalski et al.(2019)]{Kowalski2019} Kowalski, A.~F., Wisniewski, J.~P., Hawley, S.~L., et al.\ 2019, \apj, 871, 167. doi:10.3847/1538-4357/aaf058

\bibitem[Kroupa(2002)]{Kroupa2002} Kroupa, P.\ 2002, Science, 295, 82. doi:10.1126/science.1067524

\bibitem[Kulkarni et al.(2021)]{Kulkarni2021} Kulkarni, S.~R., Harrison, F.~A., Grefenstette, B.~W., et al.\ 2021, arXiv:2111.15608. doi:10.48550/arXiv.2111.15608

\bibitem[Linsky(2017)]{Linsky2017} Linsky, J.~L.\ 2017, \araa, 55, 159. doi:10.1146/annurev-astro-091916-055327

\bibitem[Linsky et al.(2020)]{Linsky2020} Linsky, J.~L., Wood, B.~E., Youngblood, A., et al.\ 2020, \apj, 902, 3. doi:10.3847/1538-4357/abb36f

\bibitem[Loyd et al.(2016)]{Loyd2016} Loyd, R.~O.~P., France, K., Youngblood, A., et al.\ 2016, \apj, 824, 102. doi:10.3847/0004-637X/824/2/102

\bibitem[Loyd et al.(2021)]{Loyd2021} Loyd, R.~O.~P., Shkolnik, E.~L., Schneider, A.~C., et al.\ 2021, \apj, 907, 91. doi:10.3847/1538-4357/abd0f0

\bibitem[Lu et al.(2019)]{Liu2019} Lu, H.-. peng ., Zhang, L.-. yun ., Shi, J., et al.\ 2019, \apjs, 243, 28. doi:10.3847/1538-4365/ab2f8f

\bibitem[Magaudda et al.(2020)]{Magaudda2020} Magaudda, E., Stelzer, B., Covey, K.~R., et al.\ 2020, \aap, 638, A20. doi:10.1051/0004-6361/201937408

\bibitem[Mann et al.(2019)]{Mann2019} Mann, A.~W., Dupuy, T., Kraus, A.~L., et al.\ 2019, \apj, 871, 63. doi:10.3847/1538-4357/aaf3bc


\bibitem[National Academies of Sciences(2021)]{Astro2020} National Academies of Sciences, E.\ 2021, Pathways to Discovery in Astronomy and Astrophysics for the 2020s, Consenses Study Report. NAtional Academies of Sciences, Engineering, and Medicine. 2021. Washington, DC: The National Academies Press, 2021.. doi:10.17226/26141

\bibitem[Newton et al.(2016)]{Newton2016} Newton, E.~R., Irwin, J., Charbonneau, D., et al.\ 2016, \apjl, 821, L19. doi:10.3847/2041-8205/821/1/L19

\bibitem[Newton et al.(2017)]{Newton2017} Newton, E.~R., Irwin, J., Charbonneau, D., et al.\ 2017, \apj, 834, 85. doi:10.3847/1538-4357/834/1/85


\bibitem[Nguyen et al.(2022)]{Nguyen2022} Nguyen, C.~T., Costa, G., Girardi, L., et al.\ 2022, \aap, 665, A126. doi:10.1051/0004-6361/202244166

\bibitem[Pal et al.(2023)]{Pal2023} Pal, T., Khan, I., Worthey, G., et al.\ 2023, \apjs, 266, 41. doi:10.3847/1538-4365/accea7

\bibitem[Parsons et al.(2018)]{Parsons2018} Parsons, S.~G., G{\"a}nsicke, B.~T., Marsh, T.~R., et al.\ 2018, \mnras, 481, 1083. doi:10.1093/mnras/sty2345

\bibitem[Paudel et al.(2024)]{Paudel2024} Paudel, R.~R., Barclay, T., Youngblood, A., et al.\ 2024, \apj, 971, 24. doi:10.3847/1538-4357/ad487d

\bibitem[Peacock et al.(2020)]{Peacock2020} Peacock, S., Barman, T., Shkolnik, E.~L., et al.\ 2020, \apj, 895, 5. doi:10.3847/1538-4357/ab893a

\bibitem[Pedregosa et al.(2020)]{scikit} Pedregosa, F., Varoquaux, G., Gramfort, A., et al. \ 2011, Journal of Machine Learning Research, 12, 2825--2830,  


\bibitem[Reinhold \& Hekker(2020)]{Reinhold2020} Reinhold, T. \& Hekker, S.\ 2020, \aap, 635, A43. doi:10.1051/0004-6361/201936887

\bibitem[Peralta et al.(2023)]{Peralta2023} Peralta, J.~I., Vieytes, M.~C., Mendez, A.~M.~P., et al.\ 2023, \aap, 676, A18. doi:10.1051/0004-6361/202346156

\bibitem[Pineda et al.(2021)]{Pineda2021} Pineda, J.~S., Youngblood, A., \& France, K.\ 2021, \apj, 918, 40. doi:10.3847/1538-4357/ac0aea

\bibitem[Rekhi et al.(2023)]{Rekhi2023} Rekhi, P., Ben-Ami, S., Perdelwitz, V., et al.\ 2023, \apj, 955, 24. doi:10.3847/1538-4357/ace5ac

\bibitem[Reiners et al.(2014)]{Reiners2014} Reiners, A., Sch{\"u}ssler, M., \& Passegger, V.~M.\ 2014, \apj, 794, 144. doi:10.1088/0004-637X/794/2/144

\bibitem[Rimmer et al.(2018)]{Rimmer2018} Rimmer, P.~B., Xu, J., Thompson, S.~J., et al.\ 2018, Science Advances, 4, eaar3302. doi:10.1126/sciadv.aar3302

\bibitem[Rugheimer et al.(2015)]{Rugheimer2015} Rugheimer, S., Segura, A., Kaltenegger, L., et al.\ 2015, \apj, 806, 137. doi:10.1088/0004-637X/806/1/137

\bibitem[Rugheimer et al.(2015)]{Rugheimer2015b} Rugheimer, S., Kaltenegger, L., Segura, A., et al.\ 2015, \apj, 809, 57. doi:10.1088/0004-637X/809/1/57

\bibitem[Saur et al.(2013)]{Saur2013} Saur, J., Grambusch, T., Duling, S., et al.\ 2013, \aap, 552, A119. doi:10.1051/0004-6361/201118179

\bibitem[Sakaue \& Shibata(2021)]{Sakaue2021} Sakaue, T. \& Shibata, K.\ 2021, \apj, 919, 29. doi:10.3847/1538-4357/ac0e34

\bibitem[Shkolnik \& Barman(2014)]{Shkolnik2014} Shkolnik, E.~L. \& Barman, T.~S.\ 2014, \aj, 148, 64. doi:10.1088/0004-6256/148/4/64

\bibitem[Schmidt et al.(2015)]{Schmidt2015} Schmidt, S.~J., Hawley, S.~L., West, A.~A., et al.\ 2015, \aj, 149, 158. doi:10.1088/0004-6256/149/5/158

\bibitem[Schneider \& Shkolnik(2018)]{Schneider2018} Schneider, A.~C. \& Shkolnik, E.~L.\ 2018, \aj, 155, 122. doi:10.3847/1538-3881/aaaa24

\bibitem[Schweitzer et al.(2019)]{Schweitzer2019} Schweitzer, A., Passegger, V.~M., Cifuentes, C., et al.\ 2019, \aap, 625, A68. doi:10.1051/0004-6361/201834965


\bibitem[Sch{\"o}fer et al.(2019)]{Schofer2019} Sch{\"o}fer, P., Jeffers, S.~V., Reiners, A., et al.\ 2019, \aap, 623, A44. doi:10.1051/0004-6361/201834114

\bibitem[Souto et al.(2020)]{Souto2020} Souto, D., Cunha, K., Smith, V.~V., et al.\ 2020, \apj, 890, 133. doi:10.3847/1538-4357/ab6d07

\bibitem[Sperauskas et al.(2016)]{Sperauskas2016} Sperauskas, J., Barta{\v{s}}i{\={u}}t{\.{e}}, S., Boyle, R.~P., et al.\ 2016, \aap, 596, A116. doi:10.1051/0004-6361/201527850

\bibitem[Stassun et al.(2019)]{Stassun2019} Stassun, K.~G., Oelkers, R.~J., Paegert, M., et al.\ 2019, \aj, 158, 138. doi:10.3847/1538-3881/ab3467

\bibitem[Tarter et al.(2007)]{Tarter2007} Tarter, J.~C., Backus, P.~R., Mancinelli, R.~L., et al.\ 2007, Astrobiology, 7, 30. doi:10.1089/ast.2006.0124

\bibitem[Taylor(2005)]{TOPCAT} Taylor, M.~B.\ 2005, Astronomical Data Analysis Software and Systems XIV, 347, 29

\bibitem[Upgren et al.(2002)]{Upgren2002} Upgren, A.~R., Sperauskas, J., \& Boyle, R.~P.\ 2002, Baltic Astronomy, 11, 91

\bibitem[Van der Walt, Colbert \& Varoquaux (2011)]{vanderWalt2011} van der Walt, S., Colbert, S.C., Varoquaux, G., \ 2011, Computing in Science
  \& Engineering, 13, 22

\bibitem[Virtanen et. al. (2020)]{Virtanen2020} Virtanen, P., Gommers, R., Oliphant, T.E., et al., \ 2020, Nature Methods, 17, 261

\bibitem[West et al.(2011)]{West2011} West, A.~A., Morgan, D.~P., Bochanski, J.~J., et al.\ 2011, \aj, 141, 97. doi:10.1088/0004-6256/141/3/97

\bibitem[Woolf \& Wallerstein(2005)]{Woolf2005} Woolf, V.~M. \& Wallerstein, G.\ 2005, \mnras, 356, 963. doi:10.1111/j.1365-2966.2004.08515.x

\bibitem[Wright et al.(2018)]{Wright2018} Wright, N.~J., Newton, E.~R., Williams, P.~K.~G., et al.\ 2018, \mnras, 479, 2351. doi:10.1093/mnras/sty1670


\bibitem[Zhang et al.(2021)]{Zhang2021} Zhang, L.-Y., Meng, G., Long, L., et al.\ 2021, \apjs, 253, 19. doi:10.3847/1538-4365/abd7a8


\end{thebibliography}
\end{document}